\documentclass{article}
\usepackage{amsmath,amsfonts,setspace,cancel,url,hyperref,verbatim,slashed,amsthm,graphicx,color,cite}

\usepackage[a4paper]{geometry}
\newcommand{\be}{\begin{equation}}
\newcommand{\ee}{\end{equation}}

\newcommand{\dd}{\mathrm{d}}		
			
\newcommand{\cH}{\mathcal{H}}			

\newcommand{\gM}{\mathcal{M}}

\newcommand{\hmu}{{\hat{\mu}}}
\newcommand{\hnu}{{\hat{\nu}}}

\newcommand{\hrho}{\hat{\rho}}
\newcommand{\hlambda}{\hat{\lambda}}
\newcommand{\hsigma}{\hat{\sigma}}

\newcommand{\1}{{\mu_1}}

\newcommand{\3}{{\mu_3}}
\newcommand{\4}{{\mu_4}}

\newcommand{\6}{{\mu_6}}
\newcommand{\7}{{\mu_7}}

\newcommand{\mt}{{\mu_{10}}}

\newcommand{\p}{\bullet}

\newcommand{\Ae}{E}

\newcommand{\Fa}{\mathcal{F}}
\newcommand{\Fb}{\mathcal{H}}
\newcommand{\Fc}{\mathcal{J}}
\newcommand{\Fd}{\mathcal{K}}
\newcommand{\Fe}{\mathcal{L}}

\numberwithin{equation}{section}

\newcommand{\G}{\mathrm{SL}(2) \times \mathbb{R}^+}
\newcommand{\Gthree}{\mathrm{SL}(3) \times \mathrm{SL}(2)}
\newcommand{\Gfour}{\mathrm{SL}(5)}
\newcommand{\Gfive}{\mathrm{SO}(5,5)}
\newcommand{\Gsix}{\mathrm{E}_6}
\newcommand{\Gseven}{\mathrm{E}_7}
\newcommand{\Geight}{\mathrm{E}_8}

\newcommand{\Hthree}{\mathrm{SO}(3) \times \mathrm{SO}(2)}
\newcommand{\Hfour}{\mathrm{SO}(5)}
\newcommand{\Hfive}{\mathrm{SO}(5)\times \mathrm{SO}(5)}
\newcommand{\Hsix}{\mathrm{USp}(8)}
\newcommand{\Hseven}{\mathrm{SU}(8)}
\newcommand{\Height}{\mathrm{SO}(16)}
\newcommand{\um}{\underline{M}} 
\newcommand{\un}{\underline{N}} 
\newcommand{\up}{\underline{P}} 
\newcommand{\uq}{\underline{Q}} 
\newcommand{\uk}{\underline{K}} 
\newcommand{\uL}{\underline{L}} 
\newcommand{\Uinv}{(U^{-1})}

\newcommand{\gV}{\mathcal{A}}
\begin{document}

\begin{titlepage}
\begin{center}

\hfill  

\vskip 1.5cm

{\Large \bf Particle actions and brane tensions from double and exceptional geometry}

\vskip 1cm

{\bf Chris D. A. Blair} \\

\vskip 25pt

{\em Theoretische Natuurkunde, Vrije Universiteit Brussel, and the International Solvay Institutes,\\
Pleinlaan 2, B-1050 Brussels, Belgium
\vskip 5pt }

{email: {\tt cblair@vub.ac.be}} \\

\end{center}

\vskip 0.5cm

\begin{center} {\bf ABSTRACT}\\[3ex]
\end{center}
\noindent
Massless particles in $n+1$ dimensions lead to massive particles in $n$ dimensions on Kaluza-Klein reduction.
In string theory, wrapped branes lead to multiplets of massive particles in $n$ dimensions, in representations of a duality group $G$.
By encoding the masses of these particles in auxiliary worldline scalars, also transforming under $G$, we write an action which resembles that for a massless particle on an extended spacetime.
We associate this extended spacetime with that appearing in double field theory and exceptional field theory, and formulate a version of the action which is invariant under the generalised diffeomorphism symmetry of these theories.
This provides a higher-dimensional perspective on the origin of mass and tension in string theory and M-theory.
Finally, we consider the reduction of exceptional field theory on a twisted torus, which is known to give the massive IIA theory of Romans. 
In this case, our particle action leads naturally to the action for a D0 brane in massive IIA. Here an extra vector field is present on the worldline, whose origin in exceptional field theory is a vector field introduced to ensure invariance under generalised diffeomorphisms.

\end{titlepage}

\newpage
\setcounter{page}{1}

\tableofcontents

\newpage
\section{Introduction}


Massless particles in $n+1$ dimensions give rise on Kaluza-Klein reduction to massive particles in $n$ dimensions. 
Consider the action for an $n+1$ dimensional massless particle: 
\be
S = \int d\tau \frac{1}{2} \lambda ( g_{\mu\nu} \dot{X}^\mu \dot{X}^\nu + \phi ( \dot{Y} + A_\mu \dot{X}^\mu )^2 )\,,
\label{toyaction1}
\ee
writing the $n+1$ dimensional metric in terms of an $n$-dimensional metric, $g_{\mu\nu}$, an $n$-dimensional vector, $A_\mu$, and an $n$-dimensional scalar $\phi$. We reduce assuming none of these fields depend on $Y$, so that it is a cyclic coordinate. We can eliminate it from the action by defining the Routhian, or partial Hamiltonian,
\be
\mathcal{H}_Y = \dot{Y} P_Y - \mathcal{L} \,,
\ee
where the momentum conjugate to $Y$ is $P_Y = \lambda \phi ( \dot{Y} + A_\mu \dot{X}^\mu )$. The Routhian is calculated to be $\mathcal{H}_Y = \frac{1}{2\lambda} \phi^{-1} P_Y^2 - \frac{\lambda}{2} g_{\mu\nu} \dot{X}^\mu \dot{X}^\nu - P_Y A_\mu \dot{X}^\mu$. If we write the action as $S = \int d\tau (\dot{Y} P_Y - \mathcal{H}_Y)$, then the $Y$ equation of motion is $\dot{P}_Y = 0$. Writing $p$ for the constant value of $P_Y$, and then integrating out the Lagrange multiplier $\lambda$, we obtain
\be
S = \int d\tau \left( -\sqrt{ \phi^{-1} p^2 }\sqrt{ - g_{\mu\nu} \dot{X}^\mu \dot{X}^\nu } + p A_\mu \dot{X}^\mu \right) \,.
\label{toyaction2}
\ee
This is a massive particle in $n$ dimensions, whose mass is given by the asymptotic value of $\sqrt{ \phi^{-1} p^2}$.

Let us run this argument backwards. Given an action of the form \eqref{toyaction2} for a massive particle in $n$ dimensions, one can encode the mass in terms of an auxiliary worldline variable, $\dot{Y}$, using an action of the form \eqref{toyaction1}. Then this action can be given a higher-dimensional interpretation.

In string theory, the above thinking is used to give the D0 brane an M-theory origin as arising from 11-dimensional momentum modes. 
Further reduction leads to more massive particle states arising from strings and branes wrapping compact cycles. On toroidal reductions, these particles will form multiplets of a duality group, $G$. 
In this paper, we will seek to understand a Kaluza-Klein-esque oxidation of these particles, where the higher-dimensional theory will appear to exist in more than 11 dimensions. The masses of the particles -- or equivalently the tensions of the branes from which they arise -- are encoded very simply in the radii of the extra dimensions.

These ideas have antecedents going back many years. 
A Kaluza-Klein origin for string and brane tensions was investigated in 
\cite{
deAzcarraga:1991px, 
Townsend:1992fa, 
Bergshoeff:1992gq}.
The idea followed is to replace the tension of a brane with a $(1+p)$-dimensional worldvolume with a dynamical $p$-form field living on the worldvolume. In the case $p=0$, for particles, there is a natural interpretation of this extra field as a higher-dimensional coordinate. This interpretation is not so clear for $p\geq 1$. However, this approach leads to some nice results. For instance, in the IIB theory, for $p=1$, the resulting tension 1-form can be combined with the worldvolume gauge field living on the D1 brane worldvolume to provide an $\mathrm{SL}(2)$ invariant description of the F1 and D1
\cite{
Townsend:1997kr, 
Cederwall:1997ts}.
This approach can be generalised to $\mathrm{SL}(2)$ invariant actions for \emph{particles} in 9 dimensions \cite{Bergshoeff:2003sy} and hence for more general $\mathrm{SL}(2)$ invariant brane actions in type IIB \cite{Bergshoeff:2006gs}.

Indeed, the starting point for the investigations described in this paper was to use the results of \cite{Bergshoeff:2003sy} for $\mathrm{SL}(2)$ invariant particles in 9 dimensions to guess the form of a general action for particles in $n$ dimensions invariant not under $\mathrm{SL}(2)$ but under some larger duality group $G$. This action is:   
\be
S =  \int d\tau\left( - \sqrt{ p_M \mathcal{M}^{MN} p_N } \sqrt{ - g_{\mu\nu} \dot{X}^\mu \dot{X}^\nu } + p_M A_\mu{}^M \dot{X}^\mu \right) \,.
\label{action2int} 
\ee
Let us explain what appears. We have a multiplet of particles transforming in a representation $R_1$ of $G$. The vector field $A_\mu{}^M$ is also in the same representation, and we have introduced charges -- or generalised momenta -- $p_M$, transforming in the representation conjugate to $R_1$. Instead of the single Kaluza-Klein scalar $\phi$ appearing in \eqref{toyaction2}, we have a set of scalars encoded in a \emph{generalised metric}, $\gM_{MN}$, which is constrained to parametrise a coset $G/H$, where $H$ is the maximal compact subgroup of the group $G$. We will check in section \ref{ndim} that this action reproduces the particle actions obtained by dimensional reduction of various brane actions exactly as expected.

To give this action a higher-dimensional interpretation, we will encode the charges $p_M$ in terms of auxiliary worldline scalars, $Y^M$. This can be done using the action
\be
S = \frac{1}{2} \int d \tau \lambda \left( g_{\mu\nu} \dot{X}^\mu \dot{X}^\nu + \gM_{MN} \left( \dot{Y}^M + \dot{X}^\mu A_\mu{}^M \right)\left( \dot{Y}^N + \dot{X}^\nu A_\nu{}^N \right)\right) \,,
\label{action1int}
\ee
where $\lambda$ is a Lagrange multiplier. 
We can treat the $Y^M$ as cyclic coordinates in a manner identical to that used above. The conjugate momenta are
\be
P_M = \lambda \mathcal{M}_{MN} ( \dot{Y}^N + A_\mu{}^N \dot{X}^\mu ) \,.
\ee
We calculate the Routhian given by Legendre transforming the Lagrangian $\mathcal{L}$ with respect to $Y^M$ but not $X^\mu$,
\be
\begin{split} 
\mathcal{H}_Y ( X^\mu, P_M) & = \dot{Y}^M P_M - \mathcal{L} \\
 & = \frac{1}{2\lambda} \mathcal{M}^{MN} P_M P_N - \frac{\lambda}{2} g_{\mu\nu} \dot{X}^\mu \dot{X}^\nu - P_M A_\mu{}^M \dot{X}^\mu \,,
\end{split}
\ee
and then trivially rewrite the action as $S = \int d\tau ( -Y^M \dot{P}_M - \mathcal{H}_Y )$. Now $Y^M$ appears only as a Lagrange multiplier enforcing the fact that $P_M$ is constant. We therefore replace $P_M = p_M$, with $p_M$ constant, so that 
\be
S = \int d\tau \left( \frac{\lambda}{2} g_{\mu\nu} \dot{X}^\mu \dot{X}^\nu - \frac{1}{2\lambda} \mathcal{M}^{MN} p_M p_N + p_M A_\mu{}^M \dot{X}^\mu \right) \,,
\ee
which after integrating out $\lambda$ corresponds to \eqref{action2int}.

The form of the action \eqref{action1int} suggests an interpretation in terms of a larger space with coordinates $(X^\mu, Y^M)$, with a metric apparently defined by $(g_{\mu\nu}, \mathcal{M}_{MN} , A_\mu{}^M)$. It would be surprising if there was a conventional higher-dimensional description, as the number of coordinates involved will be greater than 11. 

Instead, we will argue for an interpretation in terms of the structures appearing in double field theory/exceptional field theory. 
These theories are reformulations of supergravity involving the set of $G$-covariant coordinates $(X^\mu, Y^M)$, with the underlying symmetries including ``generalised diffeomorphisms'' which realise local $G$ transformations.
Recall that global $G$ is the duality group on reduction to $n$ dimensions on a $D$-torus. In double or exceptional field theory in general, one should really not call it a ``duality group'' -- duality is a statement about symmetries in certain backgrounds, such as those corresponds to toroidal reductions -- but perhaps one can refer to it here as the generalised diffeomorphism group. It plays the same role as $\mathrm{GL}(D)$ in general relativity. A key property of generalised diffeomorphisms is that they do not form a consistent algebra unless the dependence of fields and gauge parameters on the extra coordinates $Y^M$ is restricted. The simplest restriction is to impose the so-called ``section condition'', which forces one to choose a subset $Y^i$ of the $Y^M$ as the ``physical'' coordinates on which the fields of the theory can depend. 

In double field theory (DFT) \cite{Siegel:1993th, Siegel:1993xq,Hull:2009mi, Hull:2009zb, Hohm:2010jy, Hohm:2010pp} the group $G$ is $O(D,D)$. The coordinates $Y^M$ are in the fundamental of $O(D,D)$, and correspond to a doubling of a subset of (or all of) the dimensions of the original spacetime theory. In exceptional field theory, the group $G$ is $E_{D,D}$ (where $E_{D,D}$, a split real form of the exceptional groups $E_D$, is originally found as the U-duality group obtained on reducing 11-dimensional supergravity on a $D$-torus). This sequence of groups, and the $R_1$ representation of the coordinates $Y^M$, is listed in table \ref{DGH}. The development of EFT originally focused just on the subsector containing these coordinates alone \cite{Berman:2010is, Berman:2011jh,  Berman:2011cg, Berman:2012vc}, truncating the field content and the dependence on the coordinates $X^\mu$, but the full reformulation of the bosonic sector of 11-dimensional supergravity has now been carried out for every group in table \ref{DGH}, from $\G$ to $\Geight$, in 
\cite{
Berman:2015rcc, Hohm:2015xna, Musaev:2015ces, Abzalov:2015ega,Hohm:2013vpa,Hohm:2013uia,Hohm:2014fxa}. The supersymmetric versions for the $E_6$ and $E_7$ theories have also been obtained \cite{Godazgar:2014nqa, Musaev:2014lna}.

\begin{table}[h]\centering
\begin{tabular}{|c|c|c|c|c|}
\hline
$n = 11-D$ & $D$ & $R_1$ &  $G = E_{D,D}$ & $H$ \\ \hline
9 & 2 & $\mathbf{2}_1 \oplus \mathbf{2}_{-1}$ &  $\G$ & $\mathrm{SO}(2)$ \\
8 & 3 & $ (\mathbf{3,2})$ & $\Gthree$ & $\Hthree$ \\
7 & 4 & $ \mathbf{10}$ &  $\Gfour$ & $\Hfour$ \\
6 & 5 & $\mathbf{16}$ & $\Gfive$ & $\Hfive$ \\
5 & 6 & $\mathbf{27}$ & $\Gsix$ & $\Hsix$ \\
4 & 7 & $\mathbf{56}$ &  $\Gseven$ & $\Hseven$ \\
3 & 8 & $\mathbf{248}$ & $\Geight$ & $\Height$ \\
\hline
\end{tabular}
\label{DGH}
\caption{Generalised diffeomorphism groups and coordinate representations for EFT}
\end{table}

We will begin our interpretation of the action \eqref{action1int} in terms of these theories in section \ref{ndim}, where we essentially only consider a higher-dimensional space which is an extended torus. 
In section \ref{eftaction} however we will really allow all the fields to depend on the new coordinates $Y^M$. Doing so requires the introduction of an extra worldline vector transforming in the $R_1$ representation under global $G$ (but subject to some restrictions, as we will see).
This extra vector field appears to gauge the redundancy introduced by including extra coordinates, an idea that has been used in \cite{Hull:2004in, Hull:2006va} in reducing a doubled string worldsheet model to the usual string theory (similar also to the gauging procedure of \cite{Rocek:1991ps}).
It can also be seen as due to the fact that the naive ``line element'' for the extended space does not transform covariantly under the local symmetries of DFT/EFT, as was realised for DFT in \cite{Park:2013mpa, Lee:2013hma}. So this extra vector is a consequence of the fact that our local symmetries are generalised diffeomorphisms, and is fundamentally tied to the fact that this symmetry constrains the coordinate dependence of the theory through the section condition. 
Integrating out the extra coordinates and gauge fields will reduce us to particle actions in 11, 10 and $n$ dimensions. 

One could perhaps think of these dual directions as being somewhat similar to ``special isometry'' directions, such as occur in a Kaluza-Klein monopole background. 
The worldvolume action for such a brane involves an extra worldvolume vector field which gauges this isometry \cite{Bergshoeff:1997gy} and is used to eliminate what would otherwise be an extra degree of freedom corresponding to the special isometry coordinate. 

In section \ref{SSRomans}, we will point out an example where the gauge field actually survives in the reduction to 10 dimensions. This is the massive IIA supergravity of Romans \cite{Romans:1985tz}.
This is a deformation of the 10-dimensional type IIA theory which does not have a conventional 11-dimensional description.
However, it can be described within DFT and EFT in an interesting manner. In DFT one introduces a deformation by allowing the RR sector to depend linearly on a dual coordinate \cite{Hohm:2011cp}.
In EFT, the Romans deformation can be described as a deformation of the generalised diffeomorphism symmetry \cite{Ciceri:2016dmd}, which can be viewed as deriving from a generalised Scherk-Schwarz reduction of EFT 
\cite{Geissbuhler:2011mx,Aldazabal:2011nj, Grana:2012rr,Berman:2012uy, Hohm:2014qga} in which the twist matrices depend again on dual coordinates. 
The Romans supergravity can also be described in generalised geometry -- which realises $O(D,D)$ or $E_{D,D}$ symmetries on a generalised tangent bundle \cite{Gualtieri:2003dx, Hull:2007zu, Coimbra:2011nw, Coimbra:2011ky, Coimbra:2012af} -- using similar deformations of the generalised Lie derivative \cite{Cassani:2016ncu}.

Using the Scherk-Schwarz reduction procedure, our particle action gives rise to the action of a D0 brane in massive IIA, on which an extra vector field appears \cite{Bergshoeff:1997ak}. Our derivation of this fact will take a detour to highlight the fact that the EFT picture also includes the 11-dimensional non-covariant uplift of Romans supergravity described in \cite{Bergshoeff:1997ak}.

Our work hopefully sheds some light on the possible description within exceptional field theory of some parts of the brane spectrum of string theory and M-theory.
The search for ``duality covariant'' brane actions has a long history, including many papers especially relevant to the development of DFT and EFT \cite{Duff:1989tf, Duff:1990hn,Tseytlin:1990nb,Tseytlin:1990va,Siegel:1993th, Siegel:1993xq, Hull:2004in,Hull:2006va}. It has not been entirely clear how one might describe branes within EFT, where $G$ transformations relate branes of different worldvolume dimension (some other difficulties are described in \cite{ Duff:2015jka}). One attempt is \cite{Sakatani:2016sko}. 
The papers  \cite{Bandos:2015rvs, Bandos:2016ppv} study a superparticle model in which the section condition of EFT appears. 

In a sense, we are restricting ourselves to describing some aspects of the branes whose spatial worldvolumes completely wrap the internal space (and so appear as particles if we reduce to $n$ dimensions). 
These are the set of states that appear as waves -- i.e. massless particle excitations -- in the extended space, as studied as solutions of DFT/EFT in 
\cite{Berkeley:2014nza, Berman:2014jsa, Berman:2014hna} 
(see also \cite{Blair:2015eba, Park:2015bza,Naseer:2015fba} 
for the confirmation that these carry the appropriate notion of generalised momentum). The philosophy here is to think of DFT/EFT as a theory containing only massless objects, which appear as usual (massive) branes or particles on restricting to the physical spacetime.\footnote{We thank David Berman for emphasising this point to us.}


\section{Duality covariant particle actions in $n$ dimensions}
\label{ndim}

\subsection{The actions}
\label{expp} 

We repeat the two actions we wrote down in the introduction: first, the higher-dimensional form
\be
S = \frac{1}{2} \int d \tau \lambda \left( g_{\mu\nu} \dot{X}^\mu \dot{X}^\nu + \gM_{MN} \left( \dot{Y}^M + \dot{X}^\mu A_\mu{}^M \right)\left( \dot{Y}^N + \dot{X}^\nu A_\nu{}^N \right)\right) \,,
\label{action1}
\ee
which was equivalent to
\be
S =  \int d\tau\left( -\sqrt{ p_M \mathcal{M}^{MN} p_N } \sqrt{ - g_{\mu\nu} \dot{X}^\mu \dot{X}^\nu } + p_M A_\mu{}^M \dot{X}^\mu \right) \,.
\label{action2} 
\ee
We may think of these as worldline actions for particles in an $n$-dimensional spacetime. Let us repeat our description of the fields appearing. On the worldline we have scalar fields $X^\mu$ and $Y^M$. The former can be viewed as standard $n$-dimensional spacetime coordinates, while the latter will lie, as we have said, in the representation $R_1$ of the group $G$, either given by table \ref{DGH} or by $G = O(D,D)$ with $R_1 = \mathbf{2D}$.
We have an $n$-dimensional metric, $g_{\mu\nu}$, and a symmetric matrix, $\mathcal{M}_{MN}$, which parametrises a coset $G/H$, and which we refer to as the generalised metric. The vector field, $A_\mu{}^M$ also transforms in the $R_1$ representation of $G$. 
For the moment, we only allow our fields to depend on the coordinates $X^\mu$.  

To check that this action indeed corresponds to the reduction of various brane states, we should specify $n$ and $G$. 
First, let us check whether the above action corresponds to the reduction of the action for point particle states to $n$ dimensions.
We begin with the action for a massless particle in $10$ or $11$ dimensions, with metric $\hat g_{\hmu \hnu}$ and coordinates $X^{\hmu}$:
\be
S =  \int d\tau \frac{1}{2} \lambda  \hat g_{\hmu \hnu} \dot{X}^{\hmu} \dot{X}^{\hnu} 
\ee
We split $X^{\hmu} = (X^\mu,Y^i)$ and Kaluza-Klein reduce supposing the metric is independent of $Y^i$, using the decomposition
\be
\hat g_{\hmu \hnu} = 
\begin{pmatrix}  \Omega g_{\mu\nu} + \phi_{ij} A_\mu{}^i A_\nu{}^j & \phi_{ik} A_\mu{}^k \\
 \phi_{jk} A_\mu{}^k & \phi_{ij} 
\end{pmatrix} \,.
\label{gkk}
\ee
We include a conformal factor $\Omega$. 
This can be specified in order to make $g_{\mu\nu}$ either an Einstein frame metric (this is appropriate for reductions exhibiting the U-duality groups of table \ref{DGH}) or a string frame metric (appropriate for reductions exhibiting the T-duality group $O(D,D)$). 
In the latter case we have $\Omega=1$ if $\hat g_{\hmu \hnu}$ is a 10-dimensional string frame metric. In the former case, if $\hat g_{\hmu\hnu}$ is 10- or 11-dimensional Einstein frame metric then $\Omega = |\det \phi|^{-1/(n-2)}$, while if $\hat g_{\hmu\hnu}$ is the 10-dimensional string frame metric we have $\Omega = |\det \phi|^{-1/(n-2)} e^{4\Phi/(n-2)}$. 

We can eliminate the coordinates $Y^i$ in a fashion identical to the above. 
The momenta conjugate to $Y^i$ is
\be
P_i =\lambda \phi_{ij} ( \dot{Y}^j + A_\mu{}^j \dot{X}^\mu)
\ee
and the action can be written 
\be
S =  \int d\tau \left( -\sqrt{ \phi^{ij} p_i p_j } \sqrt{ - \Omega g_{\mu\nu} \dot{X}^\mu \dot X^{\nu} } + p_i A_\mu{}^i \dot{X}^\mu \right)
\label{ppred}
\ee
after setting $P_i = p_i$ constant and dropping the total derivative term $\dot{Y}^i p_i$. 

Now, returning to \eqref{action2}, we find that it matches the reduction \eqref{ppred} if 
\be
\mathcal{M}^{ij} = \Omega \phi^{ij}
\quad,\quad p_M = (p_i,0)\,.
\label{Minvij}
\ee
One can check that this agrees with the explicit form of the matrix components $\mathcal{M}^{ij}$ in all cases.\footnote{A proof in generalised geometry/DFT/EFT would note that the $\mathcal{M}^{ij}$ as a vector-vector component will only transform under the generalised Lie derivative under spacetime diffeomorphisms, and so cannot involve any $p$-form combinations. It must therefore be proportional to $\phi^{ij}$. Then one can just check the weight to confirm the $\Omega$ factor.}

Let us now check that the action \eqref{action2} corresponds to reductions of wrapped branes, and in doing so begin to comment on the relationship to double field theory and exceptional field theory. 

\subsection{$n$-dimensional particles from strings and DFT}
\label{DFTex}

\subsubsection*{Details of the $n$-dimensional theory} 

We focus now on the $n$-dimensional theory with duality group $G=O(D,D)$. Then we have coordinates $X^\mu$ and additional worldline scalars $Y^M = (Y^i,\tilde Y_i)$ transforming in the fundamental representation of $O(D,D)$. We have a metric, $g_{\mu\nu}$, and B-field, $B_{\mu\nu}$, which are invariant under $O(D,D)$, as well as a generalised metric $\gM_{MN}$ in the coset $O(D,D)/(O(D) \times O(D))$ and a one-form $A_\mu{}^M$ again in the fundamental. There is also a dilaton, which will not appear, completing the NSNS sector fields (we will not need the RR fields). 

In the double field theory \cite{Siegel:1993th, Siegel:1993xq,Hull:2009mi, Hull:2009zb, Hohm:2010jy, Hohm:2010pp} based on this $O(D,D)$, all fields depend on the coordinates $(X^\mu, Y^M)$ and transform under local $O(D,D)$ generalised diffeomorphisms (note that this corresponds to the formulation in \cite{Hohm:2013nja}, which is most similar to the set-up of exceptional field theory, with not all directions doubled). For consistency, one can impose the section condition, $\partial_i \otimes \tilde \partial^i = 0$. The canonical solution $\tilde\partial^i=0$ identifies the coordinates $Y^i$ as physical so that $(X^\mu,Y^i)$ are the genuine 10 dimensional coordinates, and the theory can be identified with (the NSNS sector of) 10-dimensional supergravity. 

We can construct a dictionary between the $O(D,D)$ covariant multiplets and the original fields $\hat g_{\hmu\hnu}$ and $\hat B_{\hmu\hnu}$ in 10 dimensions. We decompose the latter as (for the metric, this is the $\Omega=1$ case of \eqref{gkk}):
\be
\hat g_{\mu\nu} = g_{\mu\nu} + \phi_{ij} A_\mu{}^i A_\nu{}^j 
\quad , \quad
\hat g_{\mu i} = \phi_{ij} A_\mu{}^j
\quad,\quad
\hat g_{ij} = \phi_{ij} \,,
\ee
\be
\hat B_{\mu \nu} = B_{\mu\nu} - A_{[\mu}{}^j A_{\nu]j} + A_\mu{}^i A_\nu{}^j B_{ij} 
\quad,\quad
\hat B_{\mu i} = A_{\mu i} + A_\mu{}^j B_{ji}
\quad
\hat B_{ij} = B_{ij} \,.
\ee
Then the appropriate field multiplets for $O(D,D)$ are:
\be
A_\mu{}^M = \begin{pmatrix} A_\mu{}^i 
\\ A_{\mu i} 
\end{pmatrix} 
\quad,\quad
\gM_{MN} = \begin{pmatrix} \phi_{ij} - B_{ik} \phi^{kl} B_{lj} 
& B_{ik} \phi^{kj} \\
- \phi^{ik} B_{kj} & \phi^{ij} 
\end{pmatrix} \,.
\label{DFTgm}
\ee

\subsubsection*{Particles: fundamental string} 

Start with the Nambu-Goto form of the string action 
\be
S = - T \int d\tau d\sigma \left( \sqrt{ - \det \gamma_{ab}} - \hat B_2 \right)\,,
\ee
where $a,b = (\tau,\sigma)$ are worldsheet indices, the induced worldsheet metric is $\gamma_{ab} = \partial_a X^{\hmu} \partial_b X^{\hnu} \hat g_{\hmu\hnu}$, and $\hat B_2$ denotes the pullback of the $B$-field. We split the 10-dimensional coordinates $X^{\hmu}$ into $n+D$ coordinates $(X^\mu, Y^i)$ and decompose the spacetime fields as above, assuming the fields only depend on $X^\mu$. On the worldsheet, we will carry out a generalised double dimensional reduction, setting
\be
X^\mu (\tau,\sigma) = X^\mu(\tau) 
\quad ,\quad
Y^i(\tau,\sigma) = Y^i(\tau) + w^i \sigma \,.
\ee
The action is then
\be
\begin{split} 
S & = 2\pi T\int d\tau \Big(
- \sqrt{ - ( \phi_{ij} w^2 - w_i w_j ) ( \dot{Y}^i + A_\mu{}^i \dot{X}^\mu ) ( \dot{Y}^j + A_\nu{}^j \dot{X}^\nu) 
- w^2 \dot{X}^2 }
\\ & 
\qquad \qquad 
\qquad \qquad 
+ B_{ij} ( \dot{Y}^i + A_\mu{}^i \dot{X}^\mu ) w^j 
+ A_{\mu i} \dot{X}^\mu w^i 
\Big) \,.
\end{split}
\ee
We now calculate the momentum conjugate to $Y^i$, finding
\be
\frac{P_i}{2\pi T} = \frac{ ( \phi_{ij} w^2 - w_i w_j )}{\sqrt{-\det \gamma}} (\dot{Y}^j + A_\mu{}^j \dot{X}^\mu ) + B_{ij} w^j \,.
\label{ngredmom} 
\ee
By computing $P_i ( \dot{Y}^i + A_\mu{}^i \dot{X}^\mu)$ and $( P_i/2\pi T - B_{ik}w^k) \phi^{ij} ( P_j /2\pi T - B_{jl} w^l)$ we can find the Routhian. It is
\be
\begin{split} 
\mathcal{H}_Y & = \sqrt{ (P_i / 2\pi T- B_{ik} w^k) \phi^{ij} ( P_j/ 2\pi T - B_{jl} w^l ) + \phi_{ij} w^i w^j }\sqrt{ - g_{\mu\nu} \dot{X}^\mu \dot{X}^\nu }  \\  & \qquad
- \dot{X}^\mu ( A_{\mu i} w^i + A_\mu{}^i P_i /2\pi T) \,.
\end{split} 
\ee
We then use the $Y^i$ equation of motion in the action $S = \int d\tau \dot{Y}^i P_i - \mathcal{H}_Y$ to set $P_i = p_i$ to be constant. Then it is easy to see that the reduced action takes exactly the form \eqref{action2}, with the generalised metric and one-form defined in \eqref{DFTgm}, and the momenta
\be
p_M = \begin{pmatrix} p_i \\ 2\pi T w^i \end{pmatrix} \,.
\ee
For a toroidal reduction, with torus radii $R_{(i)}$, $p_i = k_i / R_{(i)}$ and $w^i = R_{(i)} m^i$, with $k_i, m_i \in \mathbb{Z}$. Then this momenta is 
\be
p_M = \begin{pmatrix} k_i/R_{(i)} \\ R_{(i)} m^i/l_s^2 \end{pmatrix} 
= \begin{pmatrix} k_i/R_{(i)} \\ m^i/\tilde R_{(i)} \end{pmatrix} 
\label{pstring}
\ee
where we have introduced the T-dual radii $\tilde R_{(i)} = l_s^2 / R_{(i)}$. We note that the momenta appearing look exactly like Kaluza-Klein momenta on a \emph{doubled} torus with radii $(R_{(i)} , \tilde R_{(i)})$. We will discuss this higher-dimensional interpretation further in section \ref{interpret}.

We must however notice that the momentum \eqref{ngredmom} obeys $w^i P_i = 0$, or $m^i k_i=0$, restricting us to have only either momenta or winding in each direction. This is a manifestation of the level-matching condition of the string. In $O(D,D)$ covariant language, we have $\eta^{MN} p_M p_N = 0$ where 
\be
\eta_{MN} = \begin{pmatrix} 0 & I \\ I & 0 \end{pmatrix}
\label{eta}
\ee
 is the defining $O(D,D)$ structure preserved by $O(D,D)$ transformations. We have not implemented this requirement in \eqref{action1}. In section \ref{psc}, we will see how some hint about how it could maybe appear when starting from a version of the action invariant under the generalised diffeomorphisms of DFT. 

\subsection{$9$-dimensional particles from branes and EFT}
\label{9dex}

\subsubsection*{Details of the $9$-dimensional theory}

We now focus on the case of maximal supergravity in 9 dimensions, which from table \ref{DGH} has a global $\G$ duality group. The representation $R_1$ of the extra worldline coordinates $Y^M$ is the reducible $\mathbf{2}_1 \oplus \mathbf{2}_{-1}$. We write $Y^M = (Y^\alpha, Y^s)$ with $\alpha=1,2$ transforming in the fundamental of $\mathrm{SL}((2)$ and $Y^s$ a singlet. 
The generalised metric, $\mathcal{M}_{MN}$, splits into a two-by-two block $\mathcal{M}_{\alpha \beta}$ and a one-by-one block, $\mathcal{M}_{ss}$. These are not independent: the determinant of $\mathcal{M}_{\alpha \beta}$ is related to $\gM_{ss}$, such that $\mathcal{H}_{\alpha \beta} = ( \gM_{ss} )^{3/4} \mathcal{M}_{\alpha \beta}$ has determinant one. 
We also have the one-form $A_\mu{}^M = (A_\mu{}^\alpha, A_\mu{}^s)$, and additional form fields which do not enter the discussion at present.

We can construct an exceptional field theory invariant under local $\G$ involving the full set of $9+3$ coordinates $(X^\mu, Y^M)$, as detailed in \cite{Berman:2015rcc}. The section condition for this theory is \cite{Wang:2015hca} $\partial_\alpha \otimes \partial_s = 0$. The solution $\partial_s \neq 0$ corresponds to IIB supergravity, so we call $Y^s$ the IIB coordinate, while $\partial_\alpha \neq 0$ corresponds to 11-dimensional supergravity. In our conventions, reduction on $Y^1$ leads to IIA supergravity in 10 dimensions, so we call $Y^1$ the M-theory direction and $Y^2$ the IIA direction.

\subsubsection*{IIA decomposition}

The 10-dimensional IIA fields are the string frame metric, $\hat g_{\hmu\hnu}$, the $B$-field, $\hat B_{\hmu\hnu}$, the dilaton $\Phi$ and the RR 1- and 3-forms, $\hat{C}_{\hmu}$ and $\hat C_{\hmu\hnu\hrho}$. 
We split the coordinates $X^{\hmu} = (X^\mu, X^9)$, identifying $X^9 \equiv Y^2$, and decompose the metric as in \eqref{gkk} with $\Omega = \phi^{-1/7} e^{4\Phi/7}$ (where $\phi \equiv |\det \phi|$). The RR 1-form is decomposed as
\be
\hat{C}_{\hmu} = \begin{pmatrix} 
C_\mu  + C_9 A_\mu \\
C_9
\end{pmatrix}\,.
\label{IIACdecomp}
\ee
Then, we have 
\be
\gM_{\alpha \beta} = \phi^{1/7} e^{10\Phi/7} \begin{pmatrix} 1 & C_9 \\ C_9 & C_9^2 + \phi e^{-2\Phi} \end{pmatrix} 
\quad,\quad
\gM^{\alpha \beta} = \phi^{-8/7} e^{4\Phi/7} \begin{pmatrix} C_9^2 + \phi e^{-2\Phi}  & - C_9 \\ - C_9 & 1 \end{pmatrix}\,,
\label{IIAgm1}
\ee
\be
\mathcal{M}_{ss} = \phi^{-6/7} e^{-4\Phi/7} \,,
\label{IIAgm2}
\ee
\be
A_\mu{}^\alpha = \begin{pmatrix} 
C_\mu \\ A_\mu
\end{pmatrix}
\quad , \quad
A_\mu{}^s = - \hat{B}_{\mu 9} \,.
\label{IIAA}
\ee
As we have identified $X^9$ with $Y^2$, note that indeed $\gM^{22} = \Omega \phi^{-1}$ as in \eqref{Minvij}.

\subsubsection*{IIA particles: fundamental string} 

Let us take $p_M = (0,0,p_s)$.
Then the action \eqref{action2} is
\be
S = |p_s| \int d\tau \left ( -\sqrt{ - \phi^{6/7} e^{4\Phi/7} g_{\mu\nu} \dot{X}^\mu \dot{X}^\nu }\pm B_{\mu 9} \dot{X}^\mu \right) \,.
\ee
This is easily seen to be the double dimensional reduction of the Nambu-Goto action for a fundamental string. 
The choice of charge vector corresponds to momentum in the IIB direction, $Y^s$, as expected.
In this setup, the string is wrapped on the $Y^2$ direction with radius $R_2$, with the T-dual IIB radius $R_s = l_s^2 / R_2$. We need to identify
\be
p_s \equiv 2\pi R_2 T_{F1} = \frac{R_2}{l_s^2} = \frac{1}{R_s} \,,
\ee 
which again exactly resembles a Kaluza-Klein momenta coming from the higher-dimensional action \eqref{action1}, as we will further discuss in section \ref{interpret}. Note that the choice of sign of $p_s$ corresponds to the orientation of the wound string. 

\subsubsection*{IIA particles: D0 brane} 

Let us take $p_M = (p_1,p_2,0)$. The action \eqref{action2} is
\be
S = \int d\tau \left ( -\sqrt{ - \phi^{-8/7} e^{4\Phi/7} ( (p_1)^2 \phi e^{-2\Phi} + (p_1 C_9 - p_2 )^2 ) g_{\mu\nu} \dot{X}^\mu \dot{X}^\nu } + ( p_1 C_\mu + p_2 A_\mu )\dot{X}^\mu \right) \,.
\label{D0action}
\ee
This is the dimensional reduction of a D0 brane carrying momentum in the direction on which we have reduced.
To see this, consider the D0 action
\be
S_{D0} = T_{D0} \int d\tau \left( -e^{-\Phi} \sqrt{ - \hat g_{\hmu\hnu} \dot{X}^\hmu \dot{X}^{\hnu} } + \dot{X}^{\hmu} \hat{C}_{\hmu} \right) \,,
\ee
and reduce using the above decomposition. We let $Z \equiv X^9$ be the direction on which we will reduce. The action is independent of $Z$ so the momentum in the $Z$ direction is conserved. This momentum is 
\be
\frac{P_Z}{T_{D0}} = \frac{\phi e^{-\Phi} ( \dot{Z} + A_\mu \dot{X}^\mu )}{\sqrt{ - g - \phi ( \dot{Z}+ A_\mu \dot{X}^\mu )^2}} + C_9 \,,
\ee
where $g \equiv \phi^{-1/7} e^{4\Phi/7} g_{\mu\nu} \dot{X}^\mu \dot{X}^\nu$. After Legendre transforming, the action can be written as
\be
S_{D0} = \int d\tau \left( - Z \dot{P}_Z - \sqrt{ - g \phi^{-1}( T_{D0}^2 \phi e^{-2\Phi} + ( P_Z - T_{D0} C_9)^2 )} +   ( T_{D0} C_\mu + P_Z A_\mu )\dot{X}^\mu \right) \,.
\ee
We solve the $Z$ equation of motion by letting $P_Z$ be constant. If the $Z$ direction has radius $R$, then let $P_Z = p/R$. Substituting back in and dropping the $\dot{Z}$ term which is now a total derivative, we find the action \eqref{D0action} with the identifications
\be
p_1 \equiv T_{D0} = \frac{1}{l_s g_s} = \frac{1}{R_1}  \quad,\quad p_2 \equiv \frac{p}{R} \,.
\ee
Here, we see that standard identification of the D0 tension with Kaluza-Klein momentum on the M-theory circle is entirely consistent with a higher-dimensional interpretation of our particle action \eqref{action1} as describing a particle moving in the extended spacetime.

\subsubsection*{IIA particles: the pp-wave} 

Finally, we take $p_M = (0,p_2,0)$ so that
\be
S = |p_2| \int d\tau \left( -\phi^{-1/2} \sqrt{ -\phi^{-1/7} e^{4\Phi/7} g_{\mu\nu} \dot{X}^\mu \dot{X}^\nu} \pm \dot{X}^\mu A_\mu \right)
\ee
This is the action for a momentum mode (compare the discussion in section \ref{expp}). It is written in terms of the lower-dimensional Einstein frame metric. Note the string frame metric in 9 dimensions would be 
$\bar{g}_{\mu\nu} = \phi^{-1/7} e^{4\Phi/7} g_{\mu\nu}$. It is trivial to identity $p_2 = p/R$ with $p \in \mathbb{Z}$. 

%

\subsubsection*{IIB decomposition} 

The bosonic fields of 10-dimensional type IIB supergravity are the Einstein frame metric, $\hat g_{\hmu \hnu}^E$, the $B$-field, $\hat B_{\hmu \hnu}$, the dilaton $\varphi$, the RR 0-form $C_0$, 2-form, $\hat C_{\hmu\hnu}$ and self-dual 4-form $\hat C_{\hmu\hnu\hrho\hat{\sigma}}$. We split the coordinates as $X^{\hmu} = (X^\mu, Y^s)$. 
We decompose the metric as in \eqref{gkk} with $\Omega = \phi^{-1/7}$. 
In the convention that $\alpha = 1$ is an RR field index and $\alpha=2$ is an NSNS field index (this is the opposite to what is \emph{stated} explicitly in \cite{Berman:2015rcc} but seems to correspond to the explicit parametrisations used there), the unit determinant part of the generalised metric can be written as
\be
\mathcal{H}_{\alpha \beta} = e^\varphi \begin{pmatrix} 1 & C_0 \\ C_0 & C_0^2 + e^{-2\varphi} \end{pmatrix} \,.
\ee
We have 
\be
{\cal M}_{ss} = \phi^{8/7} \quad,\quad
\mathcal{M}_{\alpha\beta}  = \phi^{-6/7} \mathcal{H}_{\alpha \beta} 
\quad  , \quad 
\mathcal{M}^{\alpha\beta}  = \phi^{+6/7} \mathcal{H}^{\alpha \beta} \,.
\ee
Finally, the one-form components are
\be
A_\mu{}^s = A_\mu \quad,\quad 
A_\mu{}^\alpha= \hat{C}_{\mu s}{}^\alpha \,.
\ee

\subsubsection*{IIB particles: the $pq$ string}

Take $p_M = (q_\alpha,0)$, then
\be
S = \int d\tau \left( -\sqrt{ - q_\alpha \mathcal{H}^{\alpha\beta} q_\beta \phi^{6/7} g_{\mu\nu} \dot{X}^\mu \dot{X}^\nu } + q_\alpha B_{\mu s}{}^\alpha \dot{X}^\mu \right) \,.
\ee
This matches the action for the dimensional reduction of a $pq$ string, equation (2.15) of \cite{Bergshoeff:2003sy} (excluding the Scherk-Schwarz term). We discuss the quantisation of the charges below.

\subsubsection*{IIB particles: $pp$ wave} 

Take $p_M = (0,p)$, then 
\be
S = |p| \int d\tau \left( -\phi^{-1/2}\sqrt{ - \phi^{-1/7} g_{\mu\nu} \dot{X}^\mu \dot{X}^\nu } \pm A_\mu \dot{X}^\mu \right)\,,
\ee
which is a $pp$ wave for the same reasons as above. 

\subsection{Interpretation from double and exceptional field theory} 
\label{interpret}

We have seen that the action \eqref{action2} describes $n$-dimensional particles obtained by reducing particle, string and brane actions from 10 or 11 dimensions. 
The masses of these particles are encoded in terms of the constants $p_M$, which we saw should be taken to be quantised in units of inverse radii - with the radii appearing being both the physical radii that we have reduced on and also dual radii. 
In this section, we will encode these radii in the generalised metric $\gM_{MN}$. 
Of course, this is all in accordance with standard duality relationships.
We want to emphasise in this section how this emerges from the geometry of double and exceptional field theory, given the action \eqref{action1}, so we will take the time to spell things out quite explicitly. 

The action \eqref{action1} involves what looks like the pull-back to the worldline of a ``generalised line element''
\be
``ds^2" = g_{\mu\nu} d {X}^\mu d X^\nu + \gM_{MN} (d{Y}^M + dX^\mu A_\mu{}^M)  ( d{Y}^N + dX^\nu A_\nu{}^N )\,.
\label{gle}
\ee
The Lagrange multiplier $\lambda$ then suggests to think of this action as describing massless particle-like states in an extended geometry. 

Let us focus on the particular case where the directions $Y^M$ parametrise a torus. 
We can write $\gM_{MN} dY^M dY^N = ( R_{(M)} / l)^2 \delta_{MN} dY^M dY^N$, where the dimensionful quantity $l$ can be taken as either the string length or the 11-dimensional Planck length. We denote the radius of the $Y^M$ direction by $R_{(M)}$. 
As usual, momenta in these directions should be quantised as $P_M = k_M / R_{(M)}$ where $k_M \in \mathbb{Z}$. Such momentum states will have mass, as measured using the metric $g_{\mu\nu}$, equal to $\sqrt{ \delta^{MN} P_M P_N}$.\footnote{This is the same as $\sqrt{ \gM^{MN} \bar P_M \bar P_N}$ with $\bar P_M = k_M / l_s$. Where convenient, we will in this way go back and forward between having the radii appear explicitly in the metric, or in the ranges of the coordinates.}

Let us note one can really see these standard results by applying simple particle quantum mecahnics to the action \eqref{action1}. 
The Hamiltonian is (setting $A_\mu{}^M = 0$ for simplicity here) $\mathcal{H} = g^{\mu\nu} P_\mu P_\nu + \gM^{MN} P_M P_N$, which in quantum mechanics should vanish acting on physical states. This gives an $n$-dimensional mass-shell condition $P^2 + M^2 = 0$ with $M^2 = \gM^{MN} P_M P_N$, and the usual results about quantisation of $P_M$ apply.

In this set-up, picking a solution to the section condition means selecting which $D$ of the $Y^M$ to consider as the physical coordinates. Momenta in dual directions gives rise to particles in $n$ dimensions which we would interpret ordinarily as arising from branes wrapped on the physical torus. In the action \eqref{action1}, we describe all such states as particles on the extended torus. These particles are all massless in double or exceptional field theory, as is implied by the Lagrange multiplier $\lambda$ in the action \eqref{action1}. This is consistent with the point of view of 
\cite{Berkeley:2014nza, Berman:2014jsa, Berman:2014hna}, which argued that the supergravity solutions corresponding to such totally wrapped branes appear as waves in the extended space.

We emphasise that our appproach in this paper is to take the generalised line element \eqref{gle} to be only relevant as a part of a worldline (or worldvolume) action like \eqref{action1}.
We will not think of it as corresponding to a genuine line element on the extended space (though see the paper \cite{Morand:2017fnv} which defines a metric on doubled space of DFT using an extra gauge field which can be integrated out using a path integral approach. We will meet this gauge field in the next section). 
Yet because it appears in the worldline action we can use it as proxy for inferring how point particle -- or fully wrapped brane -- states perceive the 
background of the 
doubled 
or exceptional
geometry. 

Let us confirm the generalised momenta coming from the double field theory generalised line element are what we expect. On a doubled torus we have (writing only the part of \eqref{gle} corresponding solely to the $Y^M$ directions)
\be
\begin{split} 
``ds^2" & = ( R_{(i)}/ l_s)^2 \delta_{ij} dY^i dY^j + (l_s/ R_{(i)})^2 \delta^{ij} d\tilde Y_i d\tilde Y_j \\
& = ( R_{(i)} / l_s)^2 \delta_{ij} dY^i dY^j + (\tilde R_{(i)} / l_s)^2 \delta^{ij} d\tilde Y_i d\tilde Y_j \,,
\end{split} 
\ee
so we see that this involves both the physical radii $R_{(i)}$ for the directions $Y^i$ and the dual radii $\tilde R_{(i)} = l_s^2 / R_{(i)}$ for the directions $\tilde Y_i$. The momenta appearing in the action \eqref{action2} are then $P_M = k_M / R_{(M)}$ where $k_M \in \mathbb{Z}$ and $R_{(M)} = ( R_{(i)} , \tilde R_{(i)})$. This is exactly what we saw in section \ref{DFTex} (where to be fully consistent we should there write $\phi_{ij} = \delta_{ij}$ while absorbing the radii into the definition of the coordinates $Y^M \in [0, 2\pi R_{(M)}]$). 

Now let us turn to exceptional field theory. There is a subtlety related to the fact that a conformal factor $\Omega$ appears in the dictionary relating the EFT fields to the decomposition of the 10 or 11 dimensional metric \eqref{gkk}, with $g_{\mu\nu} = \Omega^{-1}\hat g_{\mu\nu}+\dots$. We mentioned already that the inverse generalised metric has components $\gM^{ij} = \Omega \phi^{ij}$; similarly one will generically have that $\gM_{ij} = \Omega^{-1} \phi_{ij} + \dots$. This means that on an extended torus one has
\be
\mathcal{M}_{MN} dY^M dY^N =
( R_{(M)} / l)^2 \delta_{MN} dY^M dY^N 
=
\Omega^{-1} ( \tilde R_{(M)} / l)^2 \delta_{MN} dY^M dY^N 
\ee
where $\tilde R_{(M)}$ are the radii that would be seen using the 11/10 dimensional metric. These differ from the radii $R_{(M)}$ that seem to be encoded in the generalised metric, which are those seen by the metric $g_{\mu\nu}$. 
In fact, one has in general that, picking some subset $Y^i$ as the physical coordinates, 
\be
\begin{split} 
``ds^2" & = \Omega^{-1} \left( \Omega g_{\mu\nu} d {X}^\mu d X^\nu + \phi_{ij} dY^i dY^j + \dots \right)  \\
 & = \Omega^{-1} \left( \hat g_{\hmu \hnu} dX^{\hmu} dX^{\hnu} + \dots \right) \,,
\end{split} 
\label{likethat} 
\ee
where the dots denote extra terms involving both the $Y^i$ and dual coordinates. We see here the appearance of the 10/11-dimensional metric $\hat g_{\hmu \hnu}$. 

The masses measured using the metric $g_{\mu\nu}$ would be $\sqrt{ \delta^{MN} P_M P_N}$ with $P_M = k_M / R_{(M)}$ as before. We can define momenta $\tilde P_M = \Omega^{-1/2} P_M$ instead: the mass $\sqrt{\delta^{MN} \tilde P_M \tilde P_N}$ then corresponds to what would be measured using $\hat g_{\hmu\hnu}$. 

This can be viewed as a choice of redefinition of the Lagrange multiplier $\lambda$.
The freedom to redefine $\lambda$ is equivalent to rescaling both $g_{\mu\nu}$ and $\gM_{MN}$ by a conformal factor.
On choosing a parametrisation of $\gM_{MN}$ corresponding to a particular 10/11 dimensional theory, one can choose this conformal factor so that whatever radii appear correspond to those seen by the 10 or 11 dimensional metric $\hat g_{\hmu \hnu}$. 
In particular, we would define a new Lagrange multiplier $\hat \lambda = \lambda \Omega^{-1}$. 
Note that as the generalised line element is meant to only carry meaning on the worldline action, the generalised momenta defined from the action are actually unchanged:
\be
P_M = \lambda \mathcal{M}_{MN} \dot{Y}^N = \hat \lambda \Omega \mathcal{M}_{MN} \dot{Y}^N \,.
\label{sameP}
\ee
Setting $\hat \lambda = 1$ in the action \eqref{action1} corresponds to the standard results for the masses as seen in the usual 10/11 dimensional theory. This also leads to the momenta that we wrote down in section \eqref{9dex}. Ultimately, this is only really a matter of convention: we are choosing to express the masses not in terms of the $n$-dimensional metric $g_{\mu\nu}$ but in a more familiar way. 

We will now show how to use this to extract all the expected masses for particles in 9d from the $\G$ EFT. 
The results will of course be consistent with the standard duality relationships between the branes of M-theory, IIA and IIB.

In the below we drop the external metric, and write $``ds^2" = \mathcal{M}_{MN} dY^M dY^N$ only. On choosing a section, 
we explicitly extract the prefactor $\Omega^{-1}$ which will cancel against the $\hat \lambda \Omega$ in \eqref{sameP}. 
For IIA, we write
\be
``ds^2" = \phi^{1/7} e^{-4\Phi/7} \bigg( e^{2\Phi} (dY^1+ C_9 dY^2 )^2 + \phi ( dY^2 )^2 + \phi^{-1} ( dY^s)^2 \bigg) \,,
\ee
showing the prefactor $\Omega^{-1} = \phi^{1/7} e^{-4\Phi/7}$. The quantity inside the large brackets then provides what we call the ``effective radii''. 
We suppose that $\phi = (R_2/l_s)^2$, and the dilaton is constant and equal to the IIA string coupling, $e^\Phi = g_s^A$. Then we have
\be
``ds^2" = (R_2/l_s)^{2/7} (g_s^A)^{-4/7} \bigg( (g_s^A l_s/l_s)^2 (dY^1+ C_9 dY^2 )^2 + ( R_2 / l_s)^2 ( dY^2 )^2 +  (l_s / R_2)^2 ( dY^s)^2 \bigg) \,.
\label{IIAds}
\ee
The ``effective radii'' are
\be
\tilde R_s = \frac{l_s^2}{R_2}
\quad,\quad
\tilde R_1 = l_s g_s \quad,\quad
\tilde R_2 = R_2 
\ee
The momenta $p_M = k_M / \tilde R_{(M)}$ gives exactly the tensions/masses for the fundamental string wrapped on $Y^2$, the D0 brane and the pp-wave with momentum in the $Y^2$ direction.

For IIB, we have 
\be
``ds^2" = \phi^{1/7} \bigg( \phi (dY^s)^2 + \phi^{-1} e^{\varphi} \left( (dY^1 + C_0 dY^2)^2 + e^{-2\varphi} (dY^2)^2 \right) \bigg) \,.
\ee
Note that here $\phi = g_{ss}^E$ for the \emph{Einstein frame} metric. We therefore have a few extra steps to obtain results for the momenta that correspond to the masses that would be measured in the IIB string frame (we do this simply because the string frame expressions are more familiar). Letting $\phi = (R_s^E/l_s)^2$ and $e^\varphi = g_s^B$, we have
\be
``ds^2" = (R_s^E/l_s)^{2/7} \bigg( ( R_s^E/l_s)^2 (dY^s)^2 + (l_s/R_s^E)^2 g_s^B \left( (dY^1 + C_0 dY^2)^2 + (g_s^B){}^{-2}  (dY^2)^2 \right) \bigg) \,.
\ee
We have the relationship $\hat g_{\hmu \hnu}^E = e^{-\varphi/2} \hat g_{\hmu \hnu}$ for the 10-dimensional string frame $\hat g_{\hmu \hnu}$. Thus, $(R_s^E)^2 = (g_s^B)^{-1/2} (R_s)^2$. In terms of string frame quantities, we therefore have\footnote{Notice that the prefactor in both \eqref{IIAds} and \eqref{IIBds} corresponds to the T-duality invariant dilation, $e^{-2d} = e^{-2\phi} \sqrt{\det \phi}$, to the power of $2/7$.}
\be
``ds^2" = (R_s/l_s)^{2/7} (g_s^B)^{-4/7}  \bigg( ( R_s/l_s)^2 (dY^s)^2 + (l_s/R_s)^2  \left( ( g_s^B)^2 (dY^1 + C_0 dY^2)^2 +  (dY^2)^2 \right) \bigg) \,.
\label{IIBds}
\ee
The ``effective radii'' are
\be
\tilde R_s = R_s \quad ,\quad
\tilde R_1 = \frac{l_s^2g_s^B}{R_s} \quad,\quad
\tilde R_2 = \frac{l_s^2}{R_s}
\ee
The momenta $p_M = k_M /\tilde R_{(M)}$ gives exactly the tensions/masses for the pp-wave with momentum in the $Y^2$ direction, the D1 brane wrapped on $Y^s$ and the fundamental string wrapped on $Y^s$.


\section{Generalised diffeomorphism covariant particle action in extended dimensions} 
\label{eftaction}

We have already seen how the background $(g_{\mu\nu}, A_\mu{}^M, \gM_{MN})$ and coordinates $(X^\mu, Y^M)$ appearing in the action \eqref{action1} can be interpreted in terms of the fields and coordinates of double or exceptional field theory. 
So far we just considered the dictionary relating these fields to the (toroidal) reductions of brane actions to $n$ dimensions. 
In this section, we want to really interpret the action \eqref{action1} in the full DFT/EFT framework.

\subsection{Local symmetries of double and exceptional field theory}

\subsubsection*{The generalised Lie derivative}

The local symmetry transformations of these theories include ``external diffeomorphisms'', parametrised by vectors $\xi^\mu(X,Y)$, and ``generalised diffeomorphisms'', parametrised by generalised vectors, $\Lambda^M(X,Y)$. The latter realise a local infinitesimal $G$ transformation, where $G = O(D,D)$ or $E_{D,D}$. Putting DFT or EFT on a torus, global transformations of the group $G$ become the standard duality group of $n$-dimensional supergravity.

The definition of generalised diffeomorphisms $\delta_\Lambda$ (equivalently, of the generalised Lie derivative $\mathcal{L}_\Lambda$) acting on a generalised vector $V^M$ is \cite{Coimbra:2011ky,Berman:2012vc } :
\be
\delta_\Lambda V^M \equiv
\mathcal{L}_\Lambda V^M 
 = \Lambda^N \partial_N V^M 
- V^N \partial_N \Lambda^M 
+ Y^{MN}{}_{PQ} \partial_N \Lambda^P V^Q 
+ ( \lambda_V + \omega) \partial_N \Lambda^N V^M 
\label{eq:gld}
\ee 
Here $\lambda_V$ denotes the weight of the vector $V$, while we also have a sort of inherent weight $\omega$. In DFT, $\omega = 0$, while in EFT we have $\omega = -\frac{1}{n-2}$.
The tensor $Y^{MN}{}_{PQ}$ is constructed using invariants of the group $G$, and its presence ensures that the generalised Lie derivative preserves these invariants. For this to happen, the form of the $Y$-tensor is restricted and can be worked out group by group \cite{Berman:2012vc}. For $G= O(D,D)$, for instance, it is $Y^{MN}{}_{PQ} = \eta^{MN} \eta_{PQ}$ (note that in general it does not factorise in this way), while for $\G$, where the index $M= (\alpha,s)$, the non-vanishing components are $Y^{\alpha s}{}_{\beta s} = \delta^\alpha_\beta$ and those related by symmetry (it is symmetric on upper and lower indices except for the case of $E_7$). 

The gauge parameters themselves are taken to have weight $\lambda_\Lambda = - \omega$. The closure of the algebra of such transformations, 
\be
\mathcal{L}_{\Lambda_1} \mathcal{L}_{\Lambda_2} - 
\mathcal{L}_{\Lambda_2} \mathcal{L}_{\Lambda_1}
= \mathcal{L}_{[{\Lambda_1},{\Lambda_2}]_E}  
\quad,\quad [ {\Lambda_1},{\Lambda_2}]_E = \frac{1}{2} \left( \mathcal{L}_{\Lambda_1} {\Lambda_2} - \mathcal{L}_{\Lambda_2} {\Lambda_1} \right) \,,
\ee
is not guaranteed. Consistency conditions must be imposed. The simplest such condition is the section condition:
\be
Y^{MN}{}_{PQ} \partial_M \otimes \partial_N = 0 \,,
\ee
whose solutions reduce the coordinate dependence of DFT to at most 10 dimensions and that of EFT to at most 11 or 10 dimensions (there are distinct solutions giving maximal supergravity in 11 and type IIB in 10 dimensions \cite{Hohm:2013vpa, Blair:2013gqa}). 
The section condition effectively kills all dependence on the dual coordinates. Alternatively, by requiring all fields factorise in a Scherk-Schwarz (twisted) ansatz, one can find weaker conditions in which some dependence on the dual coordinates gives rise to interesting gaugings of supergravity. 

The fields $(g_{\mu\nu} , A_\mu{}^M , \gM_{MN})$ that appear in our wordline action transform as follows under generalised diffeomorphisms. The external metric $g_{\mu\nu}$ is a scalar of weight $-2\omega$. The generalised metric $\gM_{MN}$ is a tensor of zero weight. The vector field $A_\mu{}^M$ actually can be thought of as a gauge field for these transformations. Its transformation is given by
\be
\delta_\Lambda A_\mu{}^M = D_\mu \Lambda^M \equiv \partial_\mu \Lambda - \mathcal{L}_{A_\mu} \Lambda^M \,.
\ee
We take $A_\mu{}^M$ to have weight $-\omega$. The derivative $D_\mu = \partial_\mu - \mathcal{L}_{A_\mu}$ is a covariantisation of the partial derivative $\partial_\mu$ with respect to generalised diffeomorphisms. It is used in writing the action and in defining external diffeomorphisms: these are given by the usual Lie derivative with respect to parameters $\xi^\mu$, but with $\partial_\mu$ replaced by $D_\mu$. 

The field strength for $A_\mu{}^M$ is defined as follows:
\be
\mathcal{F}_{\mu \nu}{}^M  = 2 \partial_{[\mu} A_{\nu]}{}^M - [ A_\mu, A_\nu ]_E{}^M +  ( \hat\partial B_{\mu\nu} )^M \,,
\ee
in which a new two-form gauge field $B_{\mu\nu}$ appears. This field transforms in a representation of $G$ which we denote by $R_2$. (Recall that generalised vectors, and the gauge field $A_\mu{}^M$ transform in what we call $R_1$.) The derivative $\hat \partial: R_2 \rightarrow R_1$ is a nilpotent operator \cite{Cederwall:2013naa,Wang:2015hca}, constructed using group invariants and the derivatives $\partial_M$, which maps from $R_2$ to $R_1$. The representation $R_2$ is contained in the symmetric part of the tensor product $R_1 \otimes R_1$ and generally we can take
\be
( \hat \partial B_{\mu\nu} )^M = Y^{MN}{}_{PQ} \partial_N B_{\mu\nu}{}^{(PQ)} \,.
\ee
The gauge field $B_{\mu\nu}{}^{(MN)}$ has gauge transformations parametrised by one-forms $\lambda_\mu{}^{(PQ)}$, under which
\be
\delta_\lambda A_\mu{}^M = - ( \hat \partial \lambda_\mu )^M = - Y^{MN}{}_{PQ} \partial_N \lambda_\mu{}^{(PQ)} \,.
\label{Agauge}
\ee
One can go on to construct a field strength for $B_{\mu\nu}$, which necessitates the introduction of a further form field $C_{\mu\nu\rho}$, and so on leading to a ``tensor hierarchy'' (note that not all the fields that appear in this hierarchy are actually needed in the action: the point at which this occurs depends on the duality group - in $E_7$ and $E_6$ the 3-form is not used). We will not need these intricate details. 

\subsubsection*{Local symmetries including twists} 

In order to be as general as possible in specifying a particle action invariant under generalised diffeomorphisms, let us also include deformations.
This partially pre-empts some of section \ref{SSRomans}. There, we will describe how to write down a generalised Scherk-Schwarz ansatz of DFT or EFT. 
Such an ansatz involves a factorisation of the fields in terms of $Y^M$-dependent twists, which appear in the transformation rules of the fields only in certain combinations. 
We call these combinations $\Theta_{MN}{}^P$ and $\theta_M$: they must obey various consistency constraints, the first of which is that they must be constant.
These then amount to deformations of generalised diffeomorphisms. 
(The spacetime interpretation is that they provide gaugings turning supergravity into gauged supergravity - $\Theta$ is the embedding tensor, and $\theta$ is a trombone gauging.)

The precise definitions in terms of twist matrices of the Scherk-Schwarz ansatz are \eqref{theta} and \eqref{Theta}. For now, we will simply specify how they end up appearing in the symmetry transformations of our fields.
First, define a combination of these which appears naturally by
\be
\tau_{MN}{}^P = \Theta_{MN}{}^P 
+ \frac{n-2}{n-1} \left( 2 \delta^P_{[M}\theta_{N]} - Y^{PQ}{}_{MN} \theta_Q\right) \,.
\ee
The deformed generalised Lie derivative acting on a vector $V^M$ of weight $\lambda_V$ is:
\be
\begin{split} 
\delta_\Lambda V^M \equiv
\mathcal{L}_\Lambda V^M 
 &
 = \Lambda^N \partial_N V^M 
- V^N \partial_N \Lambda^M 
+ Y^{MN}{}_{PQ} \partial_N \Lambda^P V^Q 
+ ( \lambda_V + \omega) \partial_N \Lambda^N V^M 
\\ 
& 
\qquad - \tau_{NP}{}^M \Lambda^N V^P 
- \frac{\lambda_V + \omega}{\omega} \theta_N \Lambda^N V^M \,.
\end{split} 
\label{gldtwist}
\ee 
The additional gauge transformation of $A_\mu{}^M$ given in \eqref{Agauge} can also be twisted, leading to
\be
(\hat\partial \lambda_\mu)^M
= 
 Y^{MN}{}_{PQ} \partial_N \lambda_\mu{}^{(PQ)} 
- 2 \tau_{(NP)}{}^M \lambda_\mu{}^{(NP)}\,.
\label{Agaugetwist}
\ee

\subsection{The action}
\label{yetgauge}

\subsubsection*{The result} 

We now want to use the above information to think about how to write down a worldline action for a particle state coupled to the background $(g_{\mu\nu}, A_\mu{}^M, \gM_{MN})$, which respects the invariance under generalised diffeomorphisms described above. 
To do so, we need to follow 
\cite{Hull:2004in, Hull:2006va, Lee:2013hma, Ko:2016dxa} and introduce an auxiliary worldline vector field $\gV^M$, transforming in the $R_1$ representation of global $G$ (subject to the restrictions which we will come to below).
The action we find is 
\be
S = \frac{1}{2} \int d \tau \lambda \left( g_{\mu\nu} \dot{X}^\mu \dot{X}^\nu + \gM_{MN} \left( \dot{Y}^M + \gV^M + \dot{X}^\mu A_\mu{}^M \right)\left( \dot{Y}^N + \gV^N + \dot{X}^\nu A_\nu{}^N \right)\right) \,.
\label{gaction}
\ee
where under generalised diffeomorphisms \eqref{gldtwist} including twists we will require
\be
\begin{split} 
\delta_\Lambda \gV^M 
&  =
  \Lambda^P \partial_P \gV^M - \gV^P \partial_P \Lambda^M 
+ Y^{MP}{}_{KQ} \partial_P \Lambda^K ( \dot{Y}^Q + \gV^Q) 
\\ 
 & \quad - \tau_{NP}{}^M \Lambda^N (\dot{Y}^P + \gV^P) \,,
\label{deltaV} 
\end{split}
\ee
and also that the Lagrange multiplier $\lambda$ transform as a scalar with weight $+2\omega$. (This follows from the fact that the quantity in bracket naturally transforms with weight $-2\omega$, as is clear from the fact $g_{\mu\nu}$ itself does. This transformation of the Lagrange multiplier seems reminiscent of, and is perhaps ultimately inherited from, the transformation of the worldvolume metric of the M2 under duality transformations as mentioned in \cite{Duff:1990hn}. Note that for $G = O(D,D)$, $\omega = 0$.)

\subsubsection*{The reasons}

It is convenient to phrase the discussion in terms of the generalised line element:
\be
``ds^2" = g_{\mu\nu} dX^\mu dX^\nu + \gM_{MN} ( dY^M + dX^\mu A_\mu{}^M ) (dY^N + dX^\nu A_\nu{}^M )\,.
\ee
Again, we do not propose to treat this as a true metric on some extended spacetime transforming under generalised diffeomorphisms. We shall see that -- as pointed out for double field theory in \cite{Park:2013mpa} -- this quantity does not transform correctly under generalised diffeomorphisms. To remedy this, the additional field $\gV^M$ was then introduced in \cite{Lee:2013hma}.

A second motivation for introducing this gauge field is the observation \cite{Park:2013mpa} that the section condition leads to an identification of coordinates: the points $Y^M$ and 
\be
Y^M + Y^{MN}{}_{PQ} \partial_N \lambda^{(PQ)} \equiv Y^M + (\hat\partial\lambda)^M
\ee
(where $\lambda^{(PQ)}$ lives in the $R_2$ representation) may be viewed as equivalent\footnote{There is also an equivalence of generalised diffeomorphism parameters $\Lambda^M$ and $\Lambda^M + Y^{MN}{}_{PQ} \partial_N \lambda^{(PQ)}$, due to the section condition, which is a manifestation of the reducibility of $p$-form gauge transformations. The motivation for the coordinate identification is to consider some function $f(Y^M + (\hat\partial \lambda)^M) = f(Y^M) + (\hat\partial \lambda)^M \partial_M f(Y) + \dots = f(Y^M)$ after Taylor expanding and using the section condition.} and then the gauge field $\gV^M$ is introduced for this redundancy. This is akin to the gauging of \cite{Hull:2004in, Hull:2006va}, where a shift symmetry in dual directions is gauged, which is what is captured by the above equivalence. 

Our interpretation in this paper will be to treat the gauge field $\gV^M$ as an auxiliary worldline (or worldvolume) variable, which appears when writing particle (or brane) actions for DFT or EFT backgrounds. 
So we view the above ``line element'' as only having meaning on the worldline of a particle (or other brane). We mention again that one can make use of the introduction of $\gV^M$ to define a metric on the doubled space as in \cite{Morand:2017fnv}. 

The field $\gV^M$ is restricted to obey \cite{Lee:2013hma}
\be
\gV^M \partial_M = 0\,,
\ee
which is preserved by the gauge shifts $\delta_\lambda \gV^M = (\hat\partial \lambda)^M$. This means after solving the section condition, it only has components in the dual directions. As nothing depends on these directions, they are a sort of ``special isometry'' direction. Any brane in the extended space could be thought of as having such directions in addition to its usual worldvolume, transverse and special isometry directions in the physical section. Then the appearance of this vector is similar to the introducing auxiliary worldvolume vectors to gauge special isometry directions for brane action.

Possible further restrictions on $\gV^M$ will be discussed below.

\subsubsection*{The details} 

We now come to the details leading to the result \eqref{deltaV} for the transformation of $\gV$. We will consider the gauged generalised line element 
\be
\mathcal{M}_{MN}(X , Y)
( dY^{M} + \gV^{M} + d X^{\mu} A_\mu{}^M )
( dY^{N} + \gV^{N} + d X^{\nu} A_\nu{}^N )\,,
\label{gle} 
\ee
and ask how $\gV^M$ must transform for this to behave covariantly under generalised diffeomorphisms. For convenience, we will continue to write everything in terms of differentials $dY^M$ with the understanding that we really only want to consider such quantities within a worldline (or worldvolume) action, where we will replace them with worldline derivatives, $dY^M \rightarrow \dot{Y}^M$. 

Suppose we start with transformed background fields and coordinates: 
\be
\mathcal{M}_{MN}^\prime (X^\prime , Y^\prime)
( dY^{\prime M} + \gV^{\prime M} + d X^{\prime \mu} A_\mu{}^\prime{}^M )
( dY^{\prime N} + \gV^{\prime N} + d X^{\prime \nu} A_\nu{}^\prime{}^N )\,,
\label{gleprime}
\ee
where $Y^\prime = Y - \Lambda$, $X^\prime = X$.

We have
\be
\begin{split} 
\mathcal{M}_{MN}^\prime ( X, Y-\Lambda) 
& = \mathcal{M}_{MN}^\prime (X,Y) - \Lambda^P \partial_P \mathcal{M}_{MN}^\prime(X,Y) \\
& = \mathcal{M}_{MN} (X,Y) + \mathcal{L}_\Lambda \mathcal{M}_{MN}(X,Y) - \Lambda^P \partial_P \mathcal{M}_{MN} (X,Y)\,,
\end{split}
\ee
where we always work to first order in $\Lambda$. In addition,
\be
\begin{split}
A_\mu^\prime{}^M(X,Y-\Lambda) & = A_\mu^\prime{}^M(X,Y) - \Lambda^P \partial_P A_\mu^\prime{}^M (X,Y)\\ 
 & = A_\mu{}^M + D_\mu \Lambda^M - \Lambda^P \partial_P A_\mu{}^M 
-( \hat\partial \lambda_\mu)^M \,,
\end{split} 
\ee
allowing for the possibility of an extra gauge transformation which we will specify below,
\be
\begin{split} 
\gV^{\prime M} (X,Y-\Lambda) & = \gV^{\prime M}(X,Y) - \Lambda^P \partial_P \gV^{\prime M} (X,Y)
\\ 
& = \gV^M(X,Y) + \delta_\Lambda \gV(X,Y) - \Lambda^P \partial_P \gV^M(X,Y)\,,
\end{split} 
\ee
and also
\be
dY^{\prime M} = dY^M 
- dY^P \partial_P \Lambda^M - d{X}^\mu \partial_\mu \Lambda^M \,.
\ee
Note that we define the transformation under generalised diffeomorphisms by
\be
\delta_\Lambda T (Y) \equiv T^\prime(Y) - T(Y) \,,
\ee
which differs by the transport term $\Lambda^N \partial_N T(Y)$ from the total transformation $\tilde\delta_\Lambda = T^\prime(Y^\prime) - T(Y)$.

We would like, ideally, to show that the transformed expression \eqref{gleprime} equals the unprimed one \eqref{gle}. Expanding \eqref{gleprime} gives
\be
\begin{split}
\mathcal{M}_{MN}^\prime & (X^\prime , Y^\prime) 
( dY^{\prime M} + \gV^{\prime M} + d X^{\prime \mu} A_\mu{}^\prime{}^M )
( dY^{\prime N} + \gV^{\prime N} + d X^{\prime \nu} A_\nu{}^\prime{}^N )
\\ & 
= \mathcal{M}_{MN} \mathcal{D} Y^M \mathcal{D}Y^N 
+ ( \mathcal{L}_\Lambda \mathcal{M}_{MN} 
- \Lambda^P \partial_P \mathcal{M}_{MN} )
\mathcal{D} Y^M \mathcal{D}Y^N 
\\ & \qquad
+ 2 \mathcal{M}_{MN} \mathcal{D}Y^N \Big( 
- dY^P \partial_P \Lambda^M - d{X}^\mu \partial_\mu \Lambda^M 
+ \delta_\Lambda \gV^M - \Lambda^P \partial_P \gV^M 
\\& 
\qquad\qquad\qquad\qquad\qquad+ dX^\mu ( D_\mu \Lambda^M - \Lambda^P \partial_P A_\mu{}^M - ( \hat\partial \lambda_\mu)^M) 
\Big) \,.
\end{split} 
\ee
Here we abbreviated $\mathcal{D}Y^M \equiv dY^M + \gV^M + dX^\mu A_\mu{}^M$. Now, let us specify the generalised Lie derivative. We use the general form, including twists, given in \eqref{gldtwist}. Then, using $\lambda_\gM = 0$, $\lambda_\Lambda = - \omega$, we have 
\be
\begin{split}
\mathcal{L}_\Lambda \mathcal{M}_{MN} - \Lambda^P\partial_P \mathcal{M}_{MN} 
& = 2 \partial_{(M} \Lambda^P \mathcal{M}_{N)P}
- 2 Y^{PQ}{}_{K(M} \partial_Q \Lambda^K \mathcal{M}_{N)P}
- 2\omega \partial_P \Lambda^P \mathcal{M}_{MN} 
\\ & 
\qquad\qquad
\qquad\qquad
+ 2 \tau_{P(M}{}^Q \Lambda^P \mathcal{M}_{N)Q}
 + 2 \theta_P \Lambda^P \mathcal{M}_{MN} \,,
\end{split} \ee
\be
D_\mu \Lambda^M 
= \partial_\mu \Lambda^M - A_\mu{}^N \partial_N \Lambda^M
+ \Lambda^N \partial_N A_\mu{}^M 
- Y^{MN}{}_{PQ} \partial_N A_\mu{}^P \Lambda^Q 
+ \tau_{PQ}{}^M A_\mu{}^P \Lambda^Q \,.
\ee
In addition, we have the gauge transformation \eqref{Agaugetwist} of $A_\mu{}^M$.
Requiring $\mathcal{M}^\prime_{MN} \mathcal{D}Y^{\prime M} \mathcal{D}Y^{\prime N} = \mathcal{M}_{MN} \mathcal{D} Y^M \mathcal{D}Y^N$ is then equivalent to asking the following terms vanish:
\be
\begin{split}
\delta_\Lambda \gV^M &
 - \Lambda^P \partial_P \gV^M + \gV^P \partial_P \gV^M 
- Y^{MP}{}_{KQ} \partial_P \Lambda^K ( dY^Q + \gV^Q) 
\\ & + \tau_{PQ}{}^M \Lambda^P ( dY^Q + \gV^Q) 
+ (\theta_P- \omega \partial_P) \Lambda^P \mathcal{D}Y^M 
\\ & 
-Y^{MN}{}_{PQ} \partial_N ( \Lambda^{(P} A_\mu{}^{Q)} )dX^\mu 
+ 2 \tau_{PQ}{}^M \Lambda^{(P} A_\mu{}^{Q)} dX^\mu
\\ & - Y^{MN}{}_{PQ} \partial_N \lambda_\mu{}^{(PQ)} dX^\mu
+ 2 \tau_{(NP)}{}^M \lambda_\mu{}^{(NP)} dX^\mu \,.
\end{split}
\label{result}
\ee
Taking $\lambda_\mu{}^{(MN)} = \tilde \lambda_\mu{}^{(MN)} - \Lambda^{(M} A_\mu{}^{N)}$ we kill off the last two lines if 
\be
\delta_\lambda \gV^M =  dX^\mu ( Y^{MN}{}_{PQ} \partial_N \tilde \lambda_\mu{}^{(PQ)} - 2 \tau_{(NP)}{}^M \tilde \lambda_\mu{}^{(NP)} )
\label{lambdaV}
\ee
under gauge transformations \eqref{Agaugetwist} of $A_\mu{}^M$ (this means that $\mathcal{D}Y^M$ is invariant under such transformations).
We will absorb many of the remaining terms into our definition of the transformation $\delta_\Lambda \gV^M$. However, before we do so let us note that there is an issue with the weights. 
Setting the $Y$-tensor, twists, $A_\mu$ and $\gV$ to zero, we should recover ordinary differential geometry. However, in this case the unwanted terms \eqref{result} do not all vanish: an anomalous $+\omega \partial_P \Lambda^P dY^M$ term will still appear. 
This reflects the fact that the following quantity:
\be
(\sqrt{|g|})^\alpha g_{ij} dx^i dx^j \,,
\ee
where $\alpha$ is any non-zero number, is not an invariant line element. The issue is that the generalised Lie derivative is defined such that $\mathcal{M}_{MN}$ carries an intrinsic weight, while the external metric $g_{\mu\nu}$ has weight $-2\omega$.
This means that we have to relax our requirement that the quantity
\be
g_{\mu\nu} dX^\mu dX^\nu 
+ \gM_{MN} ( dY^M + \gV^M + A_\mu{}^M dX^\mu) 
(dY^N + \gV^N + A_\nu{}^N dX^\nu) 
\ee
be \emph{invariant} under generalised diffeomorphisms. Instead, it transforms as a density, provided we take the transformation 
\be
\begin{split} 
\delta_\Lambda \gV^M 
&  =
  \Lambda^P \partial_P \gV^M - \gV^P \partial_P \Lambda^M 
+ Y^{MP}{}_{KQ} \partial_P \Lambda^K ( dY^Q + \gV^Q) 
\\ 
 & \quad - \tau_{NP}{}^M \Lambda^N (dY^P + \gV^P) \,,
\label{deltaV1} 
\end{split}
\ee
which on the worldline is \eqref{deltaV}. If all we are interested in is the action \eqref{action1}, then the lack of invariance can be compensated for using the Lagrange multiplier $\lambda$, leading to the action \eqref{gaction}. We note that this means $\gV^M$ should also be taken to have the special weight $-\omega$ under generalised diffeomorphisms.

We note that term here involving the $Y$-tensor is consistent with the transformation given in \cite{Ko:2016dxa} using the condition $\gV^M \partial_M = 0$ which we have kept only in the back of our heads throughout the above calculation. We should point out that there they use the transformation $\tilde \delta$ which on the worldline is given by
\be
\begin{split} 
\tilde\delta_\Lambda Y^M & = - \Lambda^M \,, \\
\tilde\delta_\Lambda X^\mu & = 0 \,,\\
\tilde \delta_\Lambda \mathcal{O}( Y ) & = - \Lambda^P \partial_P \mathcal{O}(Y) \,,
\end{split}
\label{deltabar}
\ee
where $\mathcal{O}(Y)$ signifies any background field which depends on $Y$. The point is that though we phrased the discussion here in terms of invariance in generalised spacetime of the line element \eqref{gle}, one does not have worldline (or worldvolume more generally) invariance under generalised diffeomorphisms unless the generalised diffeomorphism corresponds to a generalised Killing vector, which annihilates the background fields. The covariance requirement on the worldline is that
\be
\tilde \delta_\Lambda ( \gM_{MN} DY^M DY^N ) 
=  - \delta_\Lambda \gM_{MN} \mathcal{D} Y^M  \mathcal{D} Y^N - 2 \gM_{MN} dX^\mu \delta_\Lambda A_\mu{}^M \mathcal{D}Y^N 
\ee
i.e. that the usual spacetime transformations $\delta_\Lambda$ of the background fields are induced. This leads to \eqref{deltaV1} without the transport term (as expected when using the transformation $\tilde \delta$).\footnote{We would like to thank A. Arvanitakis for commenting on this to us.}


\subsection{Reduction to massless particles in 10/11 dimensions}

We now study reductions of the action \eqref{gaction} corresponding to standard solutions of the section condition $Y^{MN}{}_{PQ} \partial_P \otimes \partial_Q = 0$ (this means that the extra twists $\tau_{MN}{}^P$ and $\theta_M$ can be set to zero for the remainder of this section of the paper -- they will reappear naturally in the Scherk-Schwarz reduction of section \ref{SSRomans}).

In obtaining the action \eqref{action2} from \eqref{action1}, we assumed that the fields were independent of all the extended directions $Y^M$. Now that we have figured out how to allow for field dependence on all these coordinates, subject to the section condition, we can ask what happens if we allow the fields to depend on a physical subset $Y^i$? Then the remaining coordinates - let us call them $Y^A$ - are cyclic and can easily be integrated out. The condition $\gV^M \partial_M = 0$ implies that we only have $\gV^A \neq 0$. 

To do so, we write \eqref{gaction} in the form
 \be
\begin{split}
S & = 
\int d\tau \frac{1}{2} \lambda
\Big( 
g_{\mu\nu} \dot{X}^\mu \dot{X}^\nu 
+ \left( \gM_{ij} - \gM_{iA} (\gM_{AB})^{-1} \gM_{Bj} \right)
( \dot{Y}^i + A_\mu{}^i \dot{X}^\mu ) 
( \dot{Y}^j + A_\nu{}^j \dot{X}^\nu ) 
\\ &
\qquad \qquad\qquad+ 
\gM_{AB} 
( \dot{Y}^A + \gV^A + A_\mu{}^A \dot{X}^\mu + ( \gM_{AC})^{-1} \gM_{Ci} 
( \dot{Y}^i + A_\mu{}^i \dot{X}^\mu ) )\times
\\ & \qquad\qquad\qquad\qquad\qquad\times
( \dot{Y}^B + \gV^B + A_\nu{}^B \dot{X}^\nu + ( \gM_{BD})^{-1} \gM_{Dj} 
( \dot{Y}^j + A_\nu{}^j \dot{X}^\nu ) )
\Big) \,.
\end{split} 
\ee
We consider the momenta conjugate to $Y^A$, and use the same Routhian procedure as before. 
Another result from DFT and EFT is that 
\be
\mathcal{M}_{ij} - \mathcal{M}_{i A} ( \mathcal{M}_{AB} )^{-1} \mathcal{M}_{Bj} =\Omega^{-1} \phi_{ij} \,,
\ee
while the component $A_\mu{}^i$ is identified with the vector appearing the decomposition \eqref{gkk} of the 10- or 11-dimensional metric $\hat g_{\hmu \hnu}$. 
As a result, with $\hat\lambda = \Omega^{-1} \lambda$,
\be
\begin{split}
 S & = \int d\tau \frac{1}{2} \left( \hat \lambda \hat g_{\hmu\hnu} \dot{X}^\hmu \dot{X}^\hnu - \frac{ \Omega^{-1}}{\hat{\lambda}} (\mathcal{M}_{AB})^{-1} p_A p_B \right)
 \\ & \qquad\qquad
 + p_A \int d\tau \left( \gV^A + \dot{X}^\mu A_\mu{}^A + ( \mathcal{M}_{AB})^{-1} \mathcal{M}_{Bi} ( \dot{Y}^i + \dot{X}^\nu A_\nu{}^j)\right) \,.
\label{action1011}
\end{split}
\ee
Naively, we might then integrate out $\hat\lambda$ to find the action for a particle in 10 or 11 dimensions of ``mass''
\be
M^2 = \Omega^{-1} (\mathcal{M}_{AB})^{-1} p_A p_B \,,
\ee
where the constant $p_A$, arising as the constant value of the momenta
\be
P_A = \lambda \mathcal{M}_{AB} 
( \dot{Y}^B + \gV^B + A_\mu{}^B \dot{X}^\mu + ( \gM_{BC})^{-1} \gM_{Ci} 
( \dot{Y}^i + A_\mu{}^i \dot{X}^\mu ) )\,,
\ee
appears to correspond to there being non-zero momenta in a dual direction, which one might attempt to interpret as arising from a brane winding. However, we've not made any assumptions about compact directions here, and furthermore we must not forget about the gauge field $\gV^A$. Its equation of motion set $p_A = 0$. 
Then in fact the action \eqref{action1011} becomes just that of a \emph{massless} particle in 10 or 11 dimensions:
\be
 S  = \int d\tau \frac{1}{2} \hat \lambda \hat g_{\hmu\hnu} \dot{X}^\hmu \dot{X}^\hnu \,.
\label{action10112}
\ee
The redefinition of the Lagrange multiplier is crucial here in order to match with the usual 10- or 11-dimensional metric. This redefinition of course corresponds exactly to the discussion in section \ref{yetgauge}.

We could have also integrated out $\gV^A$, or the combination $\dot{Y}^A + \gV^A$, directly, getting the same result. This is the procedure adopted in \cite{Lee:2013hma} for a doubled string action and \cite{Ko:2016dxa} for particles (where they actually start explicitly with a \emph{massive} particle in the doubled space. We prefer to begin with a massless particle in order to obtain the particle and wrapped brane states of string theory). 

\subsection{Reduction to massive particles in $n$ dimensions} 
\label{psc}

We would also like to reobtain the $n$-dimensional action \eqref{action2} from the generalised diffeomorphism invariant action \eqref{gaction}. 
Assume our background is independent of all the extended coordinates $Y^M$, so that we can integrate these out entirely.
The condition $\gV^M \partial_M = 0$ does not restrict the worldline vector $\gV^M$ at all. 
Then after integrating out we will obtain a term $\int d\tau p_M \gV^M$, and the equation of motion of $\gV^M$ then implies that $p_M = 0$, so that we can only obtain in this way a massless particle in $n$ dimensions. 

We would prefer to be able to use the action \eqref{action2} with arbitrary $p_M$. However, we see that the role of $\gV^M$ in $n$ dimensions is to kill generalised momenta in the directions in which $\gV^M$ has non-zero components. 
It is possible that there are some extra ingredients that allow us to avoid being led to $p_M = 0$. 
Firstly, we should note that we have not considered a supersymmetric form of the action \eqref{gaction}.
Secondly, we could consider restricting $\gV^M$ in different ways, by formulating constraints on $\gV^M$, which may either replace, imply or live alongside the condition $\gV^M \partial_M = 0$. 
This includes the possibility that in certain backgrounds it may be consistent to choose $\gV^M = 0$, i.e. not introduce the gauge field at all.  
We note that in general different choices of which components of $\gV^M$ are non-zero should correspond to what set of wrapped branes would exist in 10/11 dimensions, and so additional restrictions on $\gV^M$ may contain information about what branes are present. This may pertain also to topological or global information about the extended spacetime. Let us now discuss these possibilities.

\subsubsection*{Supersymmetry} 

The actions that we are studying have been solely bosonic. 
It is possible that the supersymmetric versions of \eqref{gaction} will include couplings of $\gV^M$ to fermions, so that the equation of motion of the $\gV^M$ would be modified to $p_M \neq 0$. Something similar happens in the case of the D0 brane in massive IIA, for which the bosonic action includes an extra vector field (which in section \ref{SSRomans} we will see is actually a component of $\gV^M$) whose equation of motion appears to set the Romans mass to zero. Including fermions is consistent with non-zero Romans mass \cite{Bergshoeff:1997ak}.

\subsubsection*{Restrictions on $\gV^M$}

Let us discuss possible restrictions on $\gV^M$ in more detail. 
In \cite{Ko:2016dxa, Morand:2017fnv}, the gauge field $\gV^M$ does not just obey $\gV^M\partial_M = 0$, but also is required to be null with respect to $\eta_{MN}$, the $O(D,D)$ structure: $\eta_{MN} \gV^M \gV^N = 0$. The motivation is that $\gV^M$ is the gauge field for what \cite{Park:2013mpa} called the ``coordinate gauge symmetry'' $Y^M \sim Y^M + \Delta^M$ with $\Delta^M  = \phi_1 \eta^{MN} \partial_N \phi_2$, and the gauge field is supposed to have the same behaviour as the gauge generator $\Delta^M$ which evidently satisfies $\eta_{MN} \Delta^M \Delta^N = 0$ by the section condition. 
Suppose we imposed this in the action \eqref{gaction} by a Lagrange multiplier, $\varphi$, including a term
\be
S \supset \int d\tau \left( \frac{1}{2} \varphi \eta_{MN} \gV^M \gV^N 
\right) \,.
\ee
Integrating out first $\dot{Y}^M$ leads to 
\be
S \supset \int d\tau \left( p_M \gV^M + \frac{1}{2} \varphi \eta_{MN} \gV^M \gV^N \right) \,,
\ee
and then the equation of motion for $\gV^M$ leads to
\be
S \supset -\int d\tau \frac{1}{2\varphi} \eta^{MN} p_M p_N \,.
\ee
The Lagrange multiplier $\varphi$ now restricts $p_M$ to be null with respect to $\eta$. 

Recall that in section \ref{DFTex}, we found that the generalised momenta arising from the direct dimensional reduction of the Nambu-Goto string action obeyed the condition $\eta^{MN} p_M p_N = 0$ that we impose here. The particle action \eqref{action1} was that for a massless or null particle in the doubled or extended space. If the generalised momenta are restricted to also obey the section condition, which in DFT is that they are null with respect to the $O(D,D)$ structure, we find that our actions are in a sense ``doubly null''. 

This is interesting. Does it generalise to EFT? There, we have $\Delta^M = \phi_1 Y^{MN}{}_{PQ} \partial_N\phi_2^{(PQ)}$ and it is not generally true that $Y^{MN}{}_{PQ} \Delta^P \Delta^Q = 0$. We note that in the case of DFT, the number of dual directions equals the number of physical directions. It is therefore something of an accident that one can have $\gV^M$ be null with respect to $\eta_{MN}$ and find this is compatible with enforcing the momenta also be null with respect to $\eta_{MN}$. In EFT, the condition $Y^{MN}{}_{PQ} \gV^P \gV^Q = 0$
would impose that there are the same number of non-zero components of $\gV^M$ as $\partial_M$: but this number will be less than the number of dual coordinates on picking the section $\partial_i \neq 0$, and so be more restrictive than (and generally incompatible with) $\gV^M \partial_M = 0$. 

The condition $Y^{MN}{}_{PQ} \gV^P \gV^Q = 0$ can be viewed as a ``purity condition'' on the $R_1$ valued tensor $\gV^M$ (we will explain below the reason for the terminology).
(In the language of the generalised Cartan calculus \cite{Cederwall:2013naa,Wang:2015hca} it is that the product $\gV \p \gV \in R_2$ vanishes.)
One can develop a general notion of pure $G$ tensors to describe branes in DFT/EFT \cite{CederwallKorea, MalekTalk, BCM}: given $\gV^M$ restricted as above one can formulate a differential condition defining a brane whose spatial components are wholly wrapped in the physical section. It is possible that requiring such a condition on this $\gV^M$, or on some other pure object with which $\gV^M$ must be appropriately compatible, relates to this idea. 

An approach which is similar in spirit is to use linear constraints to implement the condition $\gV^M \partial_M = 0$. This is based on \cite{Berman:2012vc}, which shows how to reformulate the section condition (a quadratic condition) as a linear condition using an auxiliary ``pure'' tensor. 
This auxiliary object $\Lambda$ transforms in some representation of $G$ and obeys a purity condition $\Lambda \otimes \Lambda |_P = 0$, where $|_P$ denotes the restriction to a particular representation (or set of representations) $P$ of $G$. The section condition can be imposed via $\Lambda \otimes \partial |_N = 0$, where $N$ is again some particular representation of $G$. 

In DFT, the section condition can be formulated in this way using a pure spinor $\Lambda$ of $O(D,D)$ (hence the terminology ``pure'' in general), satisfying $\Lambda \gamma^M \Lambda = 0$ for $\gamma^M$ the gamma matrices of $O(D,D)$. The section condition is equivalent to $\gamma^M \Lambda p_M = 0$. We note that, as we can use the $O(D,D)$ structure to raise and lower indices, that we can also require $\gV^M$ be null with respect to $\eta$ by imposing effectively the same linear constraint: $\gamma_M \Lambda \gV^M = 0$. Suppose we impose \emph{this} in the action \eqref{gaction} using a Lagrange multiplier $\varphi$ which is an $O(D,D)$ spinor. After integrating out $\dot{Y}^M$, one has the terms
\be
S \supset \int d\tau \left( p_M \gV^M + \varphi \gamma_M \Lambda \gV^M \right) \,.
\ee
The equation of motion for $\gV^M$ implies that $p_M = - \varphi \gamma_M \Lambda$. This obeys $\eta^{MN} p_M p_N = 0$ using a Fierz identity and the fact $\Lambda$ is pure; one can also show similarly that $\gamma^M \Lambda p_M = 0$. 

In appendix \ref{applin}, we show how to implement similar linear constraints for the EFT groups $G = \G$ and $G = \Gfour$. 

We note that the section condition on momenta is closely related to the BPS condition, and this may account for why it appears in this way. A particle in $n$ dimensions with arbitrary momenta $p_M$ could not be thought of as arising from the reduction of a single (BPS) brane in higher dimensions - rather, it could have momenta corresponding to e.g. M2 winding and M5 winding simultaneously. This is one physical interpretation of the condition that the generalised momenta obey the section condition. Again, everything we are doing is bosonic and it would be interesting to construct the supersymmetric version of the particle action \eqref{gaction} to learn more about these ideas.

\subsubsection*{Setting $\gV^M = 0$}

Finally, let us consider what it means in general to be able to choose $\gV^M = 0$ (which is of course one solution to the above constraints).
We are interested in backgrounds in which we can take $\partial_M = 0$. We can think of this as the most simple and extreme solution to the section condition. If so, following the general philosophy of solving the section condition means we should be applying $\partial_M = 0$ not only to our fields but also to our gauge parameters. Evidently, this is very restrictive. 
If the parameters of generalised diffeomorphisms are indeed restricted to be independent of the coordinates $Y^M$, then the action \eqref{action2} is already invariant under such transformations (which are now acting only as $X$-dependent shifts of $Y^M$ and standard gauge transformations of $A_\mu{}^M$, $\delta A_\mu{}^M = \partial_\mu \Lambda^M$). So we could argue there is no need to introduce $\gV^M$ at all. 

Let us also offer a thought about how to formalise this. 
Consider the map $\hat \partial: R_2 \rightarrow R_1$. If $B \in R_2$, then $(\hat \partial B)^M \partial_M = 0$ by the section condition.
We required $\gV^M \partial_M = 0$. We can define a map from $R_1$ to the trivial representation $\partial: R_1 \rightarrow \mathbf{1}$ by $V^M \mapsto V^M\partial_M$.
Evidently the image of $\hat \partial$ is the kernel of the latter.
One could perhaps require $\gV^M$ to be trivial in the sense that $\gV^M = (\hat \partial B)^M$ for some $B \in R_2$. 

Then, when the section is $\partial_i \neq 0$, $\partial_A = 0$, we only have components $\gV^A$ as before. However, in the section $\partial_M = 0$ in fact $\gV^M$ is zero. The action \eqref{gaction} is then identical to \eqref{action1}. More generally, one could also conceive of restricting solely to $\gV^M$ which are (equivalent to) zero in this ``cohomology''.
This may have something to do with the global or topological structure of the extended space. 

The gauge field $\gV^M$ was originally introduced in DFT in order to gauge the equivalence between $Y^M$ and $Y^M + \phi_1 \eta^{MN} \partial_N \phi_2$ due to the section condition. For $\partial_i \neq 0$, we have an equivalence $(Y^i , \tilde Y_i + \phi_1 \partial_i \phi_2)$ for arbitrary functions $\phi_{1,2}$ of the physical coordinates $Y^i$. Then one can identify all points $(Y^i, \tilde Y_i)$ and $(Y^i, \tilde Y_i + c_i)$ for arbitrary constant $c_i$ as belonging to the same gauge orbit. 

This identification of coordinates is a lot more severe than what you would want to have some notion of a genuine doubled torus (the most acceptable version of a genuinely doubled space), for which we would require only the periodic identification $(Y^i, \tilde Y_i ) \sim (Y^i, \tilde Y_i + 2\pi \tilde R_{(i)})$. 

One might suppose that introducing $\gV^M = ( 0, \tilde \gV_i )$ is what one does when one needs to gauge away entirely the dual coordinates, as perhaps would be the case when the physical spacetime is non-compact. To describe a flat doubled torus, which is a simple background in which $\partial_M = 0$, one does not introduce this gauge identification. 
However, to understand fully what is going on presumably requires a better understanding of the global properties of DFT/EFT. 

To illustrate the above points, consider the case of $\G$. We are interested in ``reducing'' the 9+3 dimensional extended space with coordinates $(X^\mu,Y^\alpha, Y^s)$ to 11 or 10 dimensions. 
(The below discussion is somewhat similar to the situation suggested presciently in \cite{AbouZeid:1999fv}.)

We claim that the section choice $\partial_\alpha \neq 0$ corresponds to a ``reduction'' on $\mathbb{R}^2 \times \{ 0 \}$. The gauge field component $\gV^s$ is non-zero and is used to gauge away the apparent dual coordinate for the (non-existent) $Y^s$ direction, equivalently, its equation of motion coming from the action \eqref{gaction} enforces that there is no momentum in this direction. 
Conversely, in the section choice $\partial_s \neq 0$, our extended spacetime is $\{ 0 \}^2 \times \mathbb{R}$. The gauge field components $\gV^\alpha$ are non-zero, and play the same role for the dual coordinates $Y^\alpha$. 

On the other, the choice $\partial_M = 0$ in which we depend on none of our coordinates can be associated to an extended space $\mathbb{T}^2 \times S^1$ (with the area of the [M-theory] torus related to the radius of the [IIB] circle). We now have $\gV^M = 0$. Our particle action now captures momentum states in all directions of the extended space. There is no standard geometrical description, meaning that there is no decompactification limit in which all three directions become non-compact. In the limit where the area of the torus goes to zero, the radius of the circle becomes infinite. The states with momentum in the circle direction can be regarded as the momentum modes of the non-compact IIB direction, while those with momentum in the torus directions become infinitely massive. The converse statements apply when the radius of the circle becomes zero, which leads to an 11-dimensional theory.


\section{Romans supergravity as EFT on a twisted torus and the D0 brane action}
\label{SSRomans}

In this final section, we will consider the effects of relaxing the section condition in order to allow some (controlled) dependence on the dual coordinates. After crossing this Rubicon, we will arrive at the Romans supergravity \cite{Romans:1985tz}. 
This is a 10-dimensional deformation of type IIA supergravity, with deformation parameter $m$ known as the Romans mass. This appears directly in the action as a sort of cosmological constant term:
\be
S_{Romans} \supset - \frac{1}{2} \int \dd^{10}X \sqrt{|\hat g|} m^2 
\ee
and appears in the gauge transformations of the form fields. Under a gauge-transformation of the B-field, $\delta \hat B_2 = d\hat\lambda_1$, we have also massive gauge transformations $\delta \hat C_1 = - m \hat \lambda_1$, $\delta \hat C_3 = - m \hat B_2 \wedge \hat \lambda_1$. The gauge invariant field strengths appearing in the action are modified due to this, with $\hat F_2 = d \hat C_1 + m \hat B_2$ and $\hat F_4 = d\hat C_3 - \hat H_3 \wedge \hat A_1 +\frac{m}{2} \hat B_2 \wedge \hat B_2$. 

The Romans supergravity is interesting within string theory, as it appears not to have a standard 11-dimensional origin. One may view it as the low energy limit of a massive IIA theory which applies in the presence of D8 branes. The Romans mass is essentially the dual of the 10-form field strength of the 9-form RR gauge field coupling to the D8. 
One can formulate a notion of ``massive T-duality'' \cite{Bergshoeff:1996ui} to relate Romans supergravity on a circle to type IIB supergravity, while also one can think of it as being related via duality to a particular compactification of M-theory on a twisted torus \cite{Hull:1998vy}.

In DFT, one can obtain the massive IIA by deforming the Ramond-Ramond sector \cite{Hohm:2011cp}, introducing a linear dependence on a dual coordinate. In EFT or generalised geometry, this deformation can be viewed as a deformation of the generalised Lie derivative \cite{Ciceri:2016dmd, Cassani:2016ncu}, which in turn can be obtained as a Scherk-Schwarz reduction of exceptional field theory on a twisted torus. The latter in particular suggests that EFT provides a higher-dimensional origin of the Romans supergravity. What is interesting is the role played by the dual coordinates in this framework. 

\subsection{Romans supergravity as a Scherk-Schwarz reduction}
\label{SSgeneral}

\subsubsection*{Scherk-Schwarz reductions of EFT}

We will largely follow \cite{Hohm:2014qga,Ciceri:2016dmd}. 
The procedure is to specify a Scherk-Schwarz or twisted ansatz for all fields of the theory. The Scherk-Schwarz twists depend on some of the coordinates $Y^M$ subject to various consistency constraints, and the fields that appear in the particle action factorise as follows:
\be
\gM_{MN} (X,Y) = U_M{}^{\um} (Y) U_N{}^{\un}(Y) \bar{\gM}_{\um \un}(X,Y) \,,
\ee
\be
e^a{}_\mu ( X,Y) = \rho^{-2\lambda}(Y) \bar{e}^a{}_\mu (X,Y) \,,
\ee
\be
A_\mu{}^M(X,Y) = \rho^{-2\lambda} (Y) \Uinv_{\um}{}^M (Y) \bar{A}_\mu{}^{\um}(X,Y) \,,
\ee
where we have written the ansatz for the vielbein of the external metric, $g_{\mu\nu} =e^a{}_\mu e^b{}_\nu \eta_{ab}$. 
We also assume that gauge parameters for generalised diffeomorphisms factorise similarly:
\be
\Lambda^M(X,Y) = \rho^{-2\lambda}(Y) \Uinv_{\um}{}^M (Y) \bar{\Lambda}^{\um}(X,Y) \,.
\ee
This can be extended to the other gauge fields of the EFT, however we will not really need these. We denote the fields that will appear in the Scherk-Schwarz reduced theory with bars on both the fields and their indices. We are being as general as possible and allowing them to still depend on some of the extended coordinates.
To do so, we have to require 
\be
\rho^{-2\lambda} (Y) \Uinv_{\um}{}^N(Y) \partial_N \bar{V} (X,Y)= \partial_{\um} \bar{V}(X,Y)\,,
\label{eq:constraint1}
\ee
i.e. the twist is trivial in directions on which the barred fields depend.

The generalised fluxes can be extracted from the transformation rules of the fields of the reduced theory. For instance, one has
\be
\begin{split} 
\delta_\Lambda e^a{}_\mu & \equiv \rho^{-2\lambda} \bar\delta_{\bar \Lambda} \bar e^a{}_\mu \\
 & =\rho^{-2\lambda} \left(  \bar\Lambda^{\um} \partial_{\um} \bar e^a{}_\mu + \lambda \partial_{\um} \bar\Lambda^{\um} \bar e^a{}_\mu 
 + \bar{\Lambda}^{\um} \theta_{\um} \bar e^a{}_\mu \right) \,,
\end{split} 
\ee
where
\be
\begin{split} 
\theta_{\um} 
 & = \frac{1}{n-2} \rho^{-2\lambda} \left( \partial_M \Uinv_{\um}{}^M - (n- 1) \Uinv_{\um}{}^M \partial_M \ln \rho^{2\lambda} \right) \,.
\end{split} 
\label{theta}
\ee
If $U^M$ and $V^M$ carry the specific weight $\lambda$, then
\be
\begin{split} 
\mathcal{L}_U V^M & \equiv \rho^{-2\lambda} \Uinv_{\um}{}^M \bar{\mathcal{L}}_{\bar U} \bar{V}^{\um} \\
 & = \rho^{-2\lambda} \Uinv_{\um}{}^M  \Big( 
\bar U^{\un} \partial_{\un} \bar V^{\um} 
- \bar V^{\un} \partial_{\un} \bar U^{\um} 
+ Y^{\um\un}{}_{\up\uq} \partial_{\un} \bar U^{\up} \bar V^{\uq} 
-\tau_{\up \uq}{}^{\um} \bar U^{\up} \bar V^{\uq} 
\Big) \,,
\end{split} 
\ee
where
\be
\tau_{\up \uq}{}^{\um} = \Theta_{\up \uq}{}^{\um} 
+ \frac{n-2}{n-1} \left( 
\delta_{\up}^{\um} \theta_{\uq} 
-\delta_{\uq}^{\um} \theta_{\up} 
- Y^{\um \un}{}_{\up\uq} \theta_{\un} \right) \,,
\label{tautheta}
\ee
with
\be
\begin{split} 
\Theta_{\up \uq}{}^{\um} 
 & = \rho^{-2\lambda} \Big(
U_K{}^{\um} \Uinv_{\uq}{}^N \partial_N \Uinv_{\up}{}^K 
- U_K{}^{\um} \Uinv_{\up}{}^N \partial_N \Uinv_{\uq}{}^K 
\\ & 
- Y^{KN}{}_{PQ} U_K{}^{\um} \Uinv_{\uq}{}^Q \partial_N \Uinv_{\up}{}^P 
\\ 
& 
- \frac{1}{n-1} \left(
\delta_{\up}^{\um} \partial_N \Uinv_{\uq}{}^N -
\delta_{\uq}^{\um} \partial_N \Uinv_{\up}{}^N 
- Y^{\um \un}{}_{\up\uq} \partial_N \Uinv_{\un}{}^N \right) 
\Big) \,.
\end{split} 
\label{Theta}
\ee
This is the embedding tensor. 

For this ansatz to make sense, various consistency conditions follow \cite{Hohm:2014qga,Ciceri:2016dmd}. These replace, and are weaker than, the section condition. For instance, we have the quadratic constraints
\be
2 \tau_{[\up| \uL}{}^{\uk} \tau_{|\uq ] \uk}{}^{\um} 
+ \tau_{\uk\uL}{}^{\um} \tau_{[\up \uq]}{}^{\uk}=0
\ee
and constraints like
\be
\tau_{\um \un}{}^{\up} \partial_{\up} \bar{V} = 0 \quad,\quad
Y^{MN}{}_{PQ} \partial_M \Uinv_{\uq}{}^Q \partial_N \bar{V} = 0\,.
\ee
In addition, the section condition should still hold on the derivatives $\partial_{\um}$ acting on the fields of the reduced theory.

\subsubsection*{$\G$ EFT on a twisted torus and Romans supergravity}

The example we will consider is to take the $\G$ EFT and reduce it on a twisted torus. 
Recall that the $R_1$ representation of this EFT was the reducible $\mathbf{2}_1 \oplus \mathbf{1}_{-1}$, and that the generalised metric $\gM_{MN}$ consisted of two blocks $\gM_{\alpha \beta}$ and $\gM_{ss}$. The unit determinant part of the former was $\mathcal{H}_{\alpha \beta} = ( \gM_{ss} )^{3/4} \mathcal{M}_{\alpha \beta}$.  
We can generically write this in terms of a complex scalar $\tau = \tau_1 + i\tau_2$,
\be
\mathcal{H}_{\alpha \beta} = \frac{1}{\tau_2} \begin{pmatrix} 1 & \tau_1 \\ \tau_1 & |\tau|^2 \end{pmatrix} \,,
\ee
which we will interpret as the complex structure of the torus (in the IIB section, this is the complex axio-dilaton, in the M-theory section on a torus, it would genuinely be the complex structure of a physical torus). We could therefore write the internal ``line element'' $\mathcal{M}_{MN} dY^M dY^N$ as 
\be
``ds^2"= ( \gM_{ss} )^{-3/4} \left( \frac{1}{\tau_2} \left( dY^1 + \tau_1 dY^2 \right)^2 + \tau_2 (dY^2 )^2 \right) 
+ \gM_{ss}^{3/4} (dY^s)^2 \,.  
\ee
The gauging which gives us the Romans supergravity is:
\be
U_{\alpha}{}^{\underline{\beta}}( Y^M ) = \begin{pmatrix} 1 & 0 \\ mY^s & 1 \end{pmatrix} \quad , \quad U_s{}^{\underline{s}} = 1\quad, \quad \rho(Y^M) = 1 \,.
\label{gauging}
\ee
We thus have
\be
\gM_{\alpha \beta} ( X, Y^M ) = U_{\alpha}{}^{\underline{\alpha}} (Y^s) U_{\beta}{}^{\underline{\beta}} (Y^s) \bar{\gM}_{\underline{\alpha} \underline\beta } ( X, Y^\alpha) \quad,\quad
\gM_{ss}(X,Y^M ) = \delta_s{}^{\underline s} \delta_s{}^{\underline s} \bar{\gM}_{\underline s \underline s} (X,Y^\alpha) \,.
\ee
The effect of the gauging is to set $\tau_1(X,Y^M) = \bar{\tau}_1(X,Y^\alpha) + m Y^s$.

The EFT background on which we are reducing can be seen to be a twisted torus by ``freezing out'' the fields of the reduced theory, i.e. setting $\bar{\gM}_{\um \un}$ to the identity. Then we see that this gauging comes from
\be
``ds^2"= \left( dY^1 + m Y^s dY^2 \right)^2 + (dY^2 )^2  
+  (dY^s)^2 \,,
\ee
which one would like to think of as a twisted torus (where owing to the restrictions on the generalised metric, there should be some relationship between the radius of the $Y^s$ direction, viewed as an $S^1$ base, and the area of the $Y^\alpha$ directions, viewed as a $T^2$ fibre). For $Y^s \rightarrow Y^s + 2\pi$, $Y^1 \rightarrow Y^1 - 2\pi m Y^2$. This is the usual coordinate patching for a twisted torus.\footnote{This is assuming the validity of giving such a precise geometric interpretation to the extended space of the EFT. At the very least though, we argue that from the point of view of the actions we are considering, particle states do ``see'' a twisted torus.}
When we carry out the Scherk-Schwarz reduction, we end up with a theory that no longer sees the $Y^s$ direction. Thus the twisted torus is only there from the point of view of the full EFT. Note that the appearance of the twisted torus here is analogous to its appearance in \cite{Hull:1998vy}.

We stress that the gauging \eqref{gauging} depends on the IIB coordinate $Y^s$. We will interpret the effective fields and gauge parameters of our reduced theory as depending on the coordinates $Y^\alpha$ of the M-theory section. In fact, from $U^{-1}_{\um}{}^{N} \partial_N \bar{V} = \partial_{\um} \bar{V}$ we see that fields and gauge parameters in the reduced theory should be taken to be independent of $Y^1$.

The above gauging induces a single non-vanishing component of the generalised fluxes: 
\be
\tau_{\underline s\, \underline 2}\,{}^{\underline 1} = \Theta_{\underline s \, \underline 2}\,{}^{\underline 1} =  m \,.
\ee
The constraints are satisfied, assuming the fields do not depend on $Y^1$. 

The appendix contains the explicit details of the action and deformations of the $\G$ EFT. Here, let us just explain a few points. The EFT action contains a ``scalar potential'' term
\be
S \supset \int \dd^{9}X\dd^3 Y \, \sqrt{|g|} V( \gM, g ) \,,
\ee
which contains all terms involving just the generalised metric, external metric and their derivatives with respect to the extended coordinates. The full expression is \eqref{eq:SSV}. 
One can show that inserting the Scherk-Schwarz ansatz for the Romans theory leads to
\be
\int \dd^9X \dd^3 Y \sqrt{ | g |} V( \gM , g) 
= \int \dd^9X \dd^3Y \sqrt{|\bar g|} \left( V ( \bar{\gM} , \bar g) 
- 
\frac{1}{2} \sqrt{|\bar g|} m^2 ( \bar{\cH}_{\underline 1\underline1} )^2 \bar{\gM}^{\underline s\underline s} 
\right) \,
\label{Vdeformed}
\ee
(where the bars again mean that these are the fields of the effective Scherk-Schwarz reduced theory). Using the relationship between the EFT fields and those of IIA, it is easy to see that new term proportional to $m^2$ is actually
\be
-\frac{1}{2} \sqrt{ |\hat g|} \, m^2 \,,
\ee
where $\hat g$ here denotes the 10-dimensional string frame metric. This is exactly the Romans mass term.

Meanwhile, the EFT gauge fields are also deformed. This is described in the appendix, and is equivalent to making the replacements
\be
\hat F_{\hmu \hnu} \rightarrow  \hat F_{\hmu \hnu} + m \hat B_{\hmu \hnu} 
\quad,\quad
\hat F_{\hmu\hnu\rho\hsigma} 
\rightarrow 
\hat F_{\hmu\hnu\rho\hsigma} + 3 m \hat B_{[\hmu \hnu} \hat B_{\hrho \hsigma]}\,.
\ee
These are exactly the modified field strengths of the Romans theory.
Using these deformations together with the fact that we know the $\G$ reduces to the action of IIA in 10 dimensions, we immediately see that this gauging indeed provides a reduction from the 12-dimensional $\G$ EFT to the 10-dimensional massive deformation of IIA. One can also check for instance that the massive gauge transformations of the Romans theory are reproduced. 

\subsection{11-dimensional interpretation of the Romans twist}
\label{SS11}

In the above procedure we let all our fields be independent of the ``M-theory direction'' $Y^1$ and interpreted our theory in the IIA section. However, at least in principle we should be able to study the deformed theory directly in 11 dimensions, with the restriction that the $Y^1$ direction must be an isometry. 

The dictionary between the metric $\hat g_{\hmu \hnu}$ of 11-dimensional supergravity and the fields of the $\G$ EFT is contained in \cite{Berman:2015rcc}. We split the coordinates $X^{\hmu} = (X^\mu, Y^\alpha)$. The EFT generalised metric is given by
\be
\mathcal{H}_{\alpha \beta} = \phi^{-1/2} \phi_{\alpha \beta} \quad,\quad
\gM_{ss} = \phi^{-6/7} \,.
\label{Mdict1} 
\ee
Here $\phi_{\alpha \beta}$ denotes the ``internal'' components of the 11-dimensional metric, $\phi_{\alpha \beta} \equiv \hat g_{\alpha \beta}$ as usual. 

Now, the Scherk-Schwarz consistency conditions tell us that our fields must be independent of $Y^1$. 
Let $k = \frac{\partial}{\partial Y^1}$ be the vector field associated to this isometry. The norm of this vector is $k^2 = \phi_{11}$. 
Then translating the Romans mass term appearing in \eqref{Vdeformed} to M-theory variables, we find that it is:
\be
- \frac{1}{2} \sqrt{|\hat g|} \, m^2 |k^2|^2 \,.
\label{11dRomans}
\ee
One can also check that the field strength of the three-form $\hat C_{\hmu\hnu\rho}$ is replaced according to
\be
\hat F_{\hmu \hnu \hrho \hsigma} 
\rightarrow 
\hat F_{\hmu\hnu\hrho\hsigma} + 3 m \hat C_{1[\hmu \hnu} \hat C_{\hrho\hsigma]1} \,.
\ee
These deformations are identical to those used in \cite{Bergshoeff:1997ak} (up to a numerical factor in the definition of $m$), where an 11-dimensional uplift of Romans supergravity was constructed. This uplift is not the usual 11-dimensional supergravity, which is well known not to reduce to Romans supergravity. The crucial feature is the presence of the Killing vector $k$: it is a theory with a built in isometry. This isometry allows the construction of the ``cosmological constant'' term \eqref{11dRomans} which does reduce to the Romans mass term in 10 dimensions. We see here that the EFT description of Romans supergravity naturally includes its uplift to this variant of 11-dimensional supergravity. It was perhaps inevitable that this had to be true, as the 11-dimensional section was still available to us (with the restriction $\partial_1 = 0$), and it would be surprising if there was some other 11-dimensional uplift of the Romans supergravity -- however it is of interest to see that this works explicitly. 

\subsection{Massive IIA particles}

\label{SSexample} 

We start with the action \eqref{gaction} and specialise to the $\G$ EFT, imposing the Scherk-Schwarz ansatz with the gauging \eqref{gauging} that leads to massive IIA.
The action can be written (omitting bars from the indices)
\be
S = \frac{1}{2} \int d \tau \lambda \left( \bar g_{\mu\nu} \dot{X}^\mu \dot{X}^\nu + \bar{\gM}_{MN} \left( \dot{Y}^M + \bar \gV^M + \dot{X}^\mu \bar A_\mu{}^M \right)\left( \dot{Y}^N + \bar \gV^N + \dot{X}^\nu \bar A_\nu{}^N \right)\right) \,.
\label{actionSS}
\ee
Here $\bar g_{\mu\nu}, \bar{\gM}_{MN}$ and $\bar A_\mu{}^M$ only depend on the coordinate $Y^2$. We have defined 
\be
\dot{Y}^M + \bar \gV^M = U_N{}^M ( \dot{Y}^N + \gV^N ) \,.
\ee
(So note that we would identify in general $\dot{Y}^{\um} = \delta^{\um}_M \dot{Y}^M$.)
For the components, we explicitly have 
$\bar{\gV}^s = \gV^s$, $\bar{\gV}^2 = \gV^2$ and
\be
\bar{\gV}^1 = \gV^1 + mY^s ( \dot{Y}^2 + \gV^2 )\,.
\ee
We have kept all the components of the gauge fields here, however the condition $\bar{\gV}^M \partial_M = 0$ (acting on barred quantities) implies that in fact $\bar\gV^2 = 0$.

The transformation rule of $\bar{\gV}^M$ follows now from the analysis of section \ref{yetgauge}, where we included the twists in the generalised Lie derivative. Alternatively, we may note that $\dot{Y}^M + \gV^M$ transforms covariantly under generalised diffeomorphisms, and so the usual twisting process applied to it leads to the correct expression \eqref{deltaV} for the transformation of $\bar{\gV}^M$.

The action \eqref{actionSS} depends only on $\dot{Y}^s$ and not $Y^s$, and so we can easily proceed to integrate out this coordinate as before. We can either use our previous results, or just do the calculation which is especially simple for $\G$. We find after Legendre transforming that
\be
S = \int d\tau \left(
\dot{Y}^s P_s + 
\frac{1}{2} \lambda ( \bar{g}_{\mu\nu} \dot{X}^\mu \dot{X}^\nu 
+ \bar{\gM}_{\alpha \beta} D_\tau Y^\alpha D_\tau Y^\beta )
- \frac{1}{2 \lambda} \bar{\gM}^{ss} P_s P_s 
+ P_s ( \gV^s + \dot{X}^\mu A_\mu{}^s )\,,
\right) 
\ee
where $P_s = \lambda \bar{\gM}_{ss} ( \dot{Y}^s + A_\mu{}^s \dot{X}^\mu)$ is the momentum in the $Y^s$ direction, and $D_\tau Y^\alpha \equiv \dot{Y}^\alpha + \bar{\gV}^\alpha + \dot{X}^\mu A_\mu{}^\alpha$ (but recall $\bar{\gV}^2 = 0$).

We now note that $P_s$ is constant by the $Y^s$ equation of motion, and zero by the $\gV^s$ equation of motion.
The action simplifies to 
\be
S = \int d\tau \frac{\lambda}{2} \left( \bar{g}_{\mu\nu} \dot{X}^\mu \dot{X}^\nu 
+ \bar{\gM}_{\alpha \beta} D_\tau Y^\alpha D_\tau Y^\beta\right)\,,
\ee
which describes a massless particle. What is this particle? We can actually interpret it in eleven dimensions. We can use the identification \eqref{Mdict1} relating the generalised metric of the $\G$ EFT to the metric components of 11-dimensional supergravity, together with the identification of the one-form doublet $A_\mu{}^\alpha$ with the Kaluza-Klein vector of the M-theory metric as in \eqref{gkk}. The caveat is that as we have carried out a Scherk-Schwarz twisting, we are not really dealing with 11-dimensional supergravity but the deformed version which reduces to massive IIA. Still, the dictionary works. Defining $\hat\lambda = \lambda |\gamma|^{1/7}$, we find the action
\be
S = \int d\tau \hat\lambda\hat g_{\hat\mu \hat\nu} D_\tau X^{\hat \mu} D_\tau X^{\hat \nu}
\ee
where $\hat g_{\hat\mu\hat\nu}$ is the 11-dimensional metric, the coordinates are $X^{\hat \mu} = (X^\mu, Y^1,Y^2)$ and $D_\tau X^{\hat \mu}= \dot{X}^{\hat \mu} + \gV^{\hat \mu}$ 
with $D_\tau X^\mu = \dot{X}^\mu$, $D_\tau Y^1 = \dot{Y}^1 + \bar{\gV}^1$, $D_\tau Y^2 = \dot{Y}^2$. 

This is the action for the ``massive M0-brane'' i.e. a massless momentum mode in the 11-dimensional deformation of supergravity which reduces to the Romans supergravity, described in \cite{Bergshoeff:1997ak}, where we are using adapted coordinates such that the Killing vector $k$ is just $\partial/\partial Y^1$. 
The dimensional reduction of the massive M0-brane then leads to the action for a massive D0 in massive IIA:
\be
S_{mD0} = T_{D0} \int d\tau \left( -e^{-\Phi} \sqrt{ - \hat g_{\hmu\hnu} \dot{X}^\hmu \dot{X}^{\hnu} } + \dot{X}^{\hmu} \hat{C}_{\hmu} + m V_\tau \right) \,,
\label{mD0}
\ee
after defining $m V_\tau \equiv \bar{\gV}^1_\tau$ following \cite{Bergshoeff:1997ak}. 
We see that the vector $\bar{\gV}^1$ becomes an additional worldline vector. The string theory interpretation is that this arises from the endpoints of strings stretching from the D0 to the background D8 brane. (The equation of motion of $V_\tau$ appears to set $m=0$, but this is only because this is just the \emph{bosonic} part of the action.)

We can therefore consider the transformation of the worldline vector $\bar{\gV}^1$ under a massive gauge transformation, which can be extracted from \eqref{deltaV} and \eqref{lambdaV}:
\be
\delta\bar{\gV}^1_\tau 
\supset 
 - m ( \bar\Lambda^s \dot{Y}^2 + ( \bar\lambda_\mu{}^{2s} + \bar\Lambda^s A_\mu{}^2 
) \dot{X}^\mu )
\ee
On the M-theory section \cite{Berman:2015rcc}, we identify $\bar\Lambda^s = \hat\lambda_{12}$ and $\bar\lambda_\mu{}^{2s} = - ( \hat\lambda_{\mu 1} + A_\mu{}^2 \hat\lambda_{12})$, where $\hat\lambda_{\hmu \hnu}$ denotes the components of the original two-form gauge parameter in eleven dimensions. Then we find that there is a contribution 
\be
\delta\bar{\gV}^1_\tau \supset m (\hat\lambda_{21} \dot{Y}^2 + \hat\lambda_{\mu 1} \dot{X}^\mu )
\ee
to the transformation of $\bar{\gV}^1$. Reducing to type IIA, we identify $\hat \lambda_{\hmu} \equiv \hat \lambda_{\hmu 1}$ as the one-form gauge parameter of the B-field. We therefore find that 
\be
\delta \bar{\gV}^1_\tau = + m \hat\lambda_\tau  
\ee
under massive gauge transformations. This is the transformation of the worldline vector of \cite{Bergshoeff:1997ak}, and ensures that \eqref{mD0} is invariant under the transformation $\delta \hat C_{\hmu} = -m \hat\lambda_{\mu}$.

We have therefore established that our action \eqref{gaction} for point particle states in the extended spacetime of EFT leads to the correct action for a D0 brane in massive IIA, on making use of the Scherk-Schwarz ansatz.  
Crucially, this would not have been possible without the extra worldline vector field $\gV^M$, whose appearance was originally due to the generalised diffeomorphism symmetry of EFT. After deforming these symmetries to obtain the massive gauge transformations of Romans supergravity, a component of the gauge field remains in the setting of the latter theory. 


\section{Discussion}

\subsection{A brief recap}

We investigated a higher-dimensional oxidation of a particle action \eqref{action2}, which described a multiplet of particle states in $n$ dimensions transforming under a duality group $G$. This uplift led to the actions \eqref{action1} and \eqref{gaction}, in which one naturally saw structures from double and exceptional field theory appearing. In particular, the action \eqref{action1} could be interpreted as a masslessness, or null, condition on a particle state in an extended spacetime. The action \eqref{gaction} showed that in order to have invariance under the local generalised diffeomorphism symmetries of DFT/EFT, one had to introduce an auxiliary vector field on the worldline, as argued in \cite{Lee:2013hma} for a doubled string action: effectively, this auxiliary vector field is used to gauge away the dual directions \cite{Hull:2004in, Hull:2006va}. Our line of thinking offers a perspective on how to describe a subset of wrapped brane states in DFT/EFT. It was interesting to see in section \ref{SSRomans} that the extra worldline field, which ordinarily would not be present in a particle or brane action, could be shown to become the extra worldline vector field that appears on a D0 brane in massive IIA \cite{Bergshoeff:1997ak}. This made use of EFT as a higher-dimensional origin for massive IIA, by Scherk-Schwarz reducing the $\G$ EFT on a twisted torus to obtain the necessary deformations to describe massive IIA as in \cite{ Ciceri:2016dmd, Cassani:2016ncu}.

\subsection{What about branes?}

We had two types of particle actions. The action \eqref{action2} corresponded directly to a massive particle in $n$ dimensions, with mass encoded in charges $p_M$. The other, the action \eqref{action1}, used extended coordinates $Y^M$ to encode the charges, and could be interpreted as the action for massless particle states in the extended spacetime of double field theory or exceptional field theory.

It would be interesting to extend these approaches to strings and branes. 
Indeed, the gauge vector $\gV^M$ was introduced in \cite{Lee:2013hma} in order to construct an action for a string in the doubled geometry of DFT. 

The generalisation to EFT should be considered. In fact, the analogue of the action \eqref{action2} in $n$ dimensions can be worked out fairly easily for the case of the $\G$ EFT. This can be done simply by reducing brane actions to 9 dimensions and using the EFT dictionary to rewrite these in terms of natural $\G$ covariant quantities. (A useful guide for what sort of action to expect is \cite{Bergshoeff:2006gs}.)

For instance, there is an $\mathrm{SL}(2)$ doublet of strings. Let us think about this in terms of (somewhat unnaturally, maybe) IIA quantities. This doublet combines the direct dimensional reduction of the D2 brane and the transverse dimensional reduction of the F1. We can find an action for this doublet by carrying out these reductions (we also integrate out the worldvolume gauge field of the D2, and dualise the worldvolume scalar on the F1 that corresponds to the coordinate $Y^2$ on which we reduce). Here, we simply state the result ($a,b$ are worldsheet indices):
\be
\begin{split} 
S & =  \int d^2 \sigma \Big(- \sqrt{  p_{\alpha s} \mathcal{M}^{\alpha \beta} \mathcal{M}^{ss} p_{\beta s}}\sqrt{ - \det ( g_{ab} + \mathcal{M}_{ss} \mathcal{F}_a{}^s \mathcal{F}_b{}^s )} 
\\ & \qquad \qquad + \frac{1}{2} \epsilon^{ab} p_{\alpha s} ( B_{ab}{}^{\alpha s} + A_{a}{}^\alpha A_b{}^s  - 2 A_{a}{}^\alpha \mathcal{F}_b{}^s ) \Big)\,,
\end{split}
\ee
where $\mathcal{F}_a{}^s = \partial_a Y^s + A_a{}^s$, with $Y^s$ an auxiliary worldsheet scalar which corresponds to the singlet coordinate of the $\G$ EFT, and the one- and two-form fields that appear are the pullbacks of the fields of the $\G$ EFT to the worldsheet. The D2 corresponds to $p_{1s} \neq 0$ and the F1 to $p_{2s} \neq 0$. The tensions are encoded in these charges as before.

Similarly, one check that the transverse reduction of the M2 action (equivalently, the D2) to 9 dimensions gives ($a,b,c$ are worldvolume indices):
\be
\begin{split} 
S = -& \int d^3 \sigma \sqrt{\frac{1}{2} p_{\alpha \beta s} \mathcal{M}^{\alpha \gamma} \mathcal{M}^{\beta \delta} \mathcal{M}^{ss} p_{\gamma \delta s}}
\sqrt{ - \det ( g_{ab} + \mathcal{M}_{\alpha \beta} F_a{}^\alpha F_b{}^\beta )} \\ & 
+ \int d^3 \sigma \frac{1}{12} p_{\alpha \beta s} \epsilon^{abc} \left(
C_{abc}{}^{\alpha \beta s} + 2 A_a{}^\alpha A_b{}^\beta A_c^s 
+ 6 ( B_{ab}{}^{\alpha s} + A_a{}^\alpha A_b{}^s ) F_c{}^\beta
+ 6 A_a{}^s F_b{}^\alpha F_c{}^\alpha 
\right)
\end{split} 
\ee 
where $p_{\alpha \beta s} = p_s \epsilon_{\alpha \beta}$, $F_a{}^\alpha = \partial_a Y^\alpha + A_a{}^\alpha$, with the $Y^\alpha$ appearing as auxiliary worldvolume scalars which can be viewed as the doublet coordinates of the $\G$ EFT, and the other fields are those of the EFT. No dualisations were carried out. 

The challenge now would be to lift these to actions describing strings and 2-branes in the 9+3 dimensional extended space of the $\G$ EFT. 
Inspired by \cite{deAzcarraga:1991px, 
Townsend:1992fa, 
Bergshoeff:1992gq,
Townsend:1997kr, 
Cederwall:1997ts}, and using the massive to massless particle analogy, the approach may perhaps involve searching for some notion of a \emph{tensionless} brane in DFT or EFT.

\subsection{Other directions} 

We saw that one could determine the masses and tensions of wrapped brane states from a simple Kaluza-Klein analysis of the ``generalised line element'' of DFT or EFT, remembering that we should only really interpret this as such as part of the worldline theory of a particle state. We only considered simple toroidal reductions here. Then, in section \ref{SSRomans}, we analysed a twisted torus reduction of EFT leading to Romans supergravity. We are currently investigating what this means in terms of the spectrum of massive IIA \cite{BBMR}. To move further away from tori, one might want to consider for instance the description of EFT on more complicated backgrounds (such as K3 as in \cite{Malek:2016vsh}) to see whether our approach captures the description of branes totally wrapping some internal manifold leading to a duality group other than the $G$ associated to toroidal reduction. With a more complete understanding of not just particles but brane actions one could go on to study physics in non-geometric backgrounds which may be more naturally described using the DFT/EFT formalisms, for instance exotic branes \cite{deBoer:2012ma} and their electric duals \cite{Blair:2016xnn}.

There was a slightly puzzle about how to treat the gauge field $\gV^M$ on reducing the generalised diffeomorphism invariant particle action \eqref{gaction} to the $n$-dimensional particle action \eqref{action2}. One could argue that choosing $\partial_M =0$ as a solution of the section condition of DFT/EFT meant that one need not introduce $\gV^M$ at all: in this case the $n$-dimensional particle could have arbitrary generalised momenta $p_M$. Alternatively, by imposing certain linear constraints on $\gV^M$, we found that the generalised momenta $p_M$ had to obey the section condition itself. This restriction on the allowed momenta may be interpreted as a statement about the origin of the $n$-dimensional particle from a single brane in higher dimensions, and be essentially a BPS condition. We saw that this condition also arose coming from the worldsheet of the fundamental string, where it is also related to level-matching. It would be interesting to further explore these relations, in particular to understand what an uplifted brane configuration whose reduction leads to generalised momenta violating the section condition would look like from the point of view of DFT/EFT. We should also explore the relationship to the work of \cite{CederwallKorea, MalekTalk, BCM} where linear constraints are used to identify branes in DFT/EFT and construct actions for such objects. This may further clarify the properties and role of $\gV^M$.

Evidently it would be beneficial to have not just bosonic actions, as presented here, but fully supersymmetric versions. Doubled string actions can be supersymmetrised \cite{ Hull:2006va, Blair:2013noa, Bandos:2015cha, Driezen:2016tnz, Park:2016sbw}, and one could explore such an extension for the particle action \eqref{gaction}. This may help clarify the general restrictions on the extra gauge field $\gV^M$ and how they relate to restrictions (especially the imposition of the section condition) on the generalised momenta $p_M$. Here there could be a link with the superparticle models of \cite{Bandos:2015rvs, Bandos:2016ppv}. 


\section*{Acknowledgements} 

I am very grateful to Jeong-Hyuck Park for discussions and to David Berman, Emanuel Malek and Felix Rudolph for both discussions and collaboration on related topics. 
I am supported by an FWO-Vlaanderen postdoctoral fellowship,
and also by the Belgian Federal Science Policy Office through the Interuniversity Attraction Pole P7/37 ``Fundamental Interactions'', and by the FWO-Vlaanderen through the project G.0207.14N, and by the Vrije Universiteit Brussel through the Strategic Research Program ``High-Energy Physics''. 

\appendix 

\section{Linear constraints on $\gV^M$ in EFT}
\label{applin}

Following \cite{Berman:2012vc,CederwallKorea}, we show to impose linear constraints on $\gV^M$ in the cases of $G = \G$ and $G = \Gfour$ which have the effect of restricting the allowed generalised momenta $p_M$ to obey the section condition. 

We start with the $\G$ EFT.
Following the prescription in \cite{Berman:2012vc} (in which $G= \G$ was not considered) we take $\Lambda \in \bar{R}_3$, which is the representation $\mathbf{1}_{-1}$. The index structure is $\Lambda_{\alpha \beta s}$ with $\alpha \beta$ antisymmetric.
We require $\Lambda \otimes \partial |_{\bar{R}_4} = 0$, where $\bar{R}_4 = \mathbf{1}_0$. This condition is $\Lambda_{\alpha \beta s} \partial_s = 0$.
The condition on $\gV^M$ is that $\gV \otimes \Lambda |_{\bar{R}_2} = 0$, or $\Lambda_{\alpha \beta s} \gV^\beta = 0$. This implies that $\gV^\alpha = 0$. 
In the action, this leads after integrating out $\dot{Y}^M$ to
\be
S \supset \int d\tau \left( \varphi^\alpha \Lambda_{\alpha \beta s} \gV^\beta + p_\alpha \gV^\alpha + p_s \gV^s \right) \,,
\ee
from which we find $p_s = 0$ and $p_\alpha = - \varphi^\beta \Lambda_{\beta \alpha s} \neq 0$. 

Evidently, this imposes that we have no momenta in the IIB direction, $Y^s$. Equivalently, the linear condition used here only enforces the solution $\partial_s = 0$ of the section condition. It turns out that in EFT one needs a different linear constraint to give the IIB section $\partial_s \neq 0$, as was explained in \cite{CederwallKorea}. The pure object $\Lambda$ must now be taken to belong to the $R_1$ representation.
For $\G$, this is the $\mathbf{2}_1 \oplus \mathbf{1}_{-1}$. The purity condition is that we only have components in the $\mathbf{3}_2 \oplus \mathbf{1}_{-2}$ representation in the tensor product $R_1 \otimes R_1$. This means $\Lambda^s = 0$ and $\Lambda^\alpha \neq 0$. One can take $\Lambda^1 = 1$ and $\Lambda^2 =0$ as a representative. Then we impose the condition $\Lambda \otimes \partial |_{adj} = 0$. The projection into the adjoint here means that we require $\Lambda^\alpha \partial_\beta - \frac{1}{2} \delta^\alpha_\beta \Lambda^\gamma \partial_\gamma = 0$. For $\Lambda^\alpha \neq 0$ this implies $\partial_\alpha = 0$. 
We then also require $\Lambda \otimes \gV |_{R_2} = 0$, which is $\Lambda^\alpha \gV^s + \Lambda^s \gV^\alpha = \Lambda^\alpha \gV^s = 0$. This means that $\gV^s = 0$. This is in accord with the IIB section, $\partial_s \neq 0$. 
The Lagrange multiplier terms in the action then give
\be
S \supset \int d\tau \left( \varphi_{\alpha s} \Lambda^\alpha \gV^s + p_\alpha \gV^\alpha + p_s \gV^s \right) \,.
\ee
This gives $p_\alpha = 0$ and $p_s = - \varphi_{\alpha s} \Lambda^\alpha \neq 0$. 

Let's take another example, this time $G = \mathrm{SL}(5)$. 
The coordinate representation $R_1$ is the $\mathbf{10}$. We let $a$ be a five-dimensional index in the fundamental representation, so that we can write $\gV^M = \gV^{ab}$ with $ab$ antisymmetric. The other representations relevant to us are $R_2 = \mathbf{\bar 5}$, $R_3 = \mathbf{5}$ and $R_4 = \mathbf{\bar{10}}$. 

The linear constraint for the section condition solution corresponding to 11-dimensional supergravity is \cite{Berman:2012vc} $\Lambda_{[a } \partial_{bc]} = 0$. Here $\Lambda_a \in \bar{R}_3 = \mathbf{\bar 5}$. No purity condition is required. We also impose $\Lambda_b \gV^{ab} = 0$. Taking only $\Lambda_{5} \neq 0$, for example, gives $\partial_{i5} \neq 0$, $\partial_{ij} = 0$ for $i,j = 1,2,3,4$, and also corresponds to $\gV^{ij} \neq 0$, $\gV^{i5} = 0$, as we expect. In the reduction of the action \eqref{gaction}, we find
\be
S \supset \int d\tau \left( \varphi_a \Lambda_b \gV^{ab} + \frac{1}{2} p_{ab} \gV^{ab} \right) \,,
\ee
which implies $p_{ab} = - 2 \varphi_{[a} \Lambda_{b]}$, so that $\Lambda_{[a} p_{bc]} = 0$ and hence $\epsilon^{abcde} p_{ab} p_{cd} = 0$, which is the section condition for this EFT \cite{Berman:2011cg}. 

Meanwhile, the linear constraint relevant to the IIB section solution is \cite{CederwallKorea} $\Lambda^{ab} \partial_{bc} = 0$ for $\Lambda^{ab} \in R_1$ obeying $\Lambda^{[ab} \Lambda^{cd]} = 0$. We also require $\Lambda^{[ab} \gV^{cd]} = 0$. A representative solution is $\Lambda^{45} \neq 0$, which means only $\partial_{12}, \partial_{13}, \partial_{23}$ are non-zero, which is the IIB section solution \cite{Blair:2013gqa}. This also implies that $\gV^{12}, \gV^{13}$ and $\gV^{23}$ are zero and the rest non-zero, as necessary. In the action we find
\be
S \supset \int d\tau \left( \frac{1}{4} \varphi^e \epsilon_{eabcd} \Lambda^{ab} \gV^{cd} + \frac{1}{2} p_{ab} \gV^{ab} \right) \,.
\ee
This gives $p_{ab} = - \frac{1}{2} \varphi^e \epsilon_{eabcd} \Lambda^{cd}$, which implies $\Lambda^{ab} p_{bc} = 0$ and hence $\epsilon^{abcde}p_{ab} p_{cd} = 0$. 

We therefore see that one can formulate certain constraints on the gauge field $\gV^M$, which correspond to imposing the section condition on the generalised momenta which appear as charges in the $n$-dimensional action \eqref{action2}. In this case, with $\partial_M = 0$, one has access to the duality symmetry $G$ which allows one to transform any particular generalised momenta into any other in its orbit.

\section{Further details of the Scherk-Schwarz reduction of the $\G$ EFT}
\label{app}

We record in this appendix some general expressions for the Scherk-Schwarz reduction of the $\G$ EFT, which were worked out in a prior incarnation of this paper, and which may prove to have some use.

\subsection{The action}
\label{accact}

The bosonic fields of the $\G$ EFT that we encountered in the main body of this paper were the external metric, $g_{\mu\nu}$, the one-form $A_\mu{}^M$ and the generalised metric $\gM_{MN}$. The extended coordinates $Y^M$ were in the $\mathbf{2}_1 \oplus \mathbf{1}_{-1}$ of $\G$. 
In addition, there are other form fields in the tensor hierarchy. We have an $\mathrm{SL}(2)$ doublet of two-forms, $B_{\mu\nu}{}^{\alpha s}$, with field strength $\Fb_{\mu\nu\rho}{}^{\alpha s}$, a singlet three-form $C_{\mu\nu\rho}{}^{\alpha \beta s}$ (the indices $\alpha \beta$ are antisymmetric) with field strength $\Fc_{\mu\nu\rho\sigma}{}^{\alpha \beta s}$, a singlet four-form, $D_{\mu\nu\rho\sigma}{}^{\alpha \beta ss}$ with field strength $\Fd_{\mu\nu\rho\sigma\lambda}{}^{\alpha \beta ss}$, and a doublet of five-forms, $\Ae_{\mu\nu\rho\sigma\lambda\kappa}{}^{\gamma, \alpha \beta ss}$ with field strength $\Fe_{\mu\nu\rho\sigma\lambda\kappa \tau}{}^{\gamma,\alpha\beta ss}$. 
The precise definitions of these field strengths, and the gauge transformations of the gauge fields, can be found in \cite{Berman:2015rcc}. 

From the point of view of supergravity, these gauge fields encode the degrees of freedom of the various supergravity gauge fields plus their duals (so in the M-theory case, just the three-form field and its six-form dual). Hence they do not all represent independent degrees of freedom: in the $\G$ EFT, one only has kinetic terms for $A_\mu{}^M$, $B_{\mu\nu}{}^{\alpha s}$ and $C_{\mu \nu \rho}{}^{\alpha \beta s}$. 

The action is
\be
\begin{split} 
S = \int \dd^{9} X \dd^3 Y \sqrt{|g|} & \Big( 
R
-\frac{7}{32} g^{\mu \nu} D_\mu \ln \gM_{ss} D_\nu \ln \gM_{ss} 
+ \frac{1}{4}  g^{\mu \nu} D_\mu \cH_{\alpha \beta} D_\nu \cH^{\alpha \beta}
\\ & 
- \frac{1}{2\cdot 2!} \gM_{MN} \mathcal{F}_{\mu \nu}{}^M \mathcal{F}^{\mu \nu N} 
 - \frac{1}{2\cdot 3!} \gM_{\alpha \beta} \gM_{ss} \mathcal{H}_{\mu \nu \rho}{}^{\alpha s} \mathcal{H}^{\mu \nu \rho \beta s} 
\\ & - \frac{1}{2\cdot 2! 4!} \gM_{ss} \gM_{\alpha \gamma} \gM_{\beta \delta} \mathcal{J}_{\mu \nu \rho \sigma}{}^{[\alpha \beta] s} \mathcal{J}^{\mu \nu \rho \sigma [\gamma \delta]s} 
\\ & 
+ V(\mathcal{M}_{MN}, g) + \sqrt{g}^{-1} \mathcal{L}_{top} \Big) 
\end{split} 
\label{eq:S} 
\ee
where $V(\mathcal{M}_{MN},g)$ denotes the would-be scalar potential (we correct here a numerical error in the coefficients of \cite{Berman:2015rcc}) 
\be
\begin{split}
V & = 
\frac{1}{4} \gM^{ss}\left( 
 \partial_s \cH^{\alpha \beta} \partial_s \cH_{\alpha \beta} 
 + \partial_s g^{\mu \nu} \partial_s g_{\mu \nu} + \partial_s \ln g \partial_s \ln g 
 \right) 
 \\ & 
 +  \frac{9}{32}  \gM^{ss} \partial_s \ln \gM_{ss} \partial_s \ln \gM_{ss}
- \frac{1}{2}  \gM^{ss} \partial_s \ln \gM_{ss} \partial_s \ln g 
\\ &  +
\gM_{ss}^{3/4} \Bigg[ 
\frac{1}{4} \cH^{\alpha \beta} \partial_\alpha \cH^{\gamma \delta} \partial_\beta \cH_{\gamma \delta} 
- \frac{1}{2} \cH^{\alpha \beta} \partial_\alpha \cH^{\gamma \delta} \partial_\gamma \cH_{\delta \beta} 
+ \partial_\alpha \cH^{\alpha \beta} \partial_\beta \ln \left(  g^{1/2} \gM_{ss}^{3/4}  \right)  
\\ & 
\qquad\qquad + \frac{1}{4} \cH^{\alpha \beta} \left( \partial_\alpha g^{\mu \nu} \partial_\beta g_{\mu \nu} 
+   \partial_\alpha \ln g \partial_\beta \ln g
+ \frac{1}{4} \partial_\alpha \ln \gM_{ss} \partial_\beta \ln \gM_{ss}  + \frac{3}{2} \partial_\alpha \ln g \partial_\beta \ln \gM_{ss} 
 \right) \!
\Bigg] \,,
\end{split}
\ee
and the topological term may be defined most conveniently as an integral over one dimension higher as $\int d^{10}x d^3Y \tilde{\mathcal{L}}_{top}$ with
\begin{equation}
 \begin{split}
\tilde{\mathcal{L}}_{top} = \frac{1}{5!48} \varepsilon^{\1\ldots\mt}  &\frac{1}{4}  \epsilon_{\alpha \beta} \epsilon_{\gamma \delta}  \left[ 
  \frac{1}{5}  \partial_s \Fd_{\mu_1 \dots \mu_5}{}^{\alpha \beta ss} \Fd_{\mu_6 \dots \mu_{10}}{}^{\gamma \delta ss} 
 - \frac{5}{2} \Fa_{\mu_1 \mu_2}{}^s \Fc_{\3\ldots\6}{}^{\alpha \beta s} \Fc_{\7\ldots\mt}{}^{\gamma \delta s}  \right. \\
    & \qquad\qquad\qquad \left. + \frac{20}{3} \Fb_{\mu_1 \dots \mu_3}{}^{\alpha s}\Fb_{\4\ldots\6}{}^{\beta s} \Fc_{\7 \ldots \mt}{}^{\gamma \delta s}  \right] \,.
 \end{split} \label{eq:ToptTerm}
\end{equation}

\subsection{Scherk-Schwarz reduction} 
\label{appSS}

Let us now consider Scherk-Schwarz reductions of the $\G$ exceptional field theory. Owing to the reducibility of the extended coordinate representation, the twist matrix $U_M{}^{\um}$ in this case consists of doublet and singlet pieces, $U_{\alpha}{}^{\underline{\alpha}}$ and $U_{s}{}^{\underline{s}}$. Here $U_\alpha{}^{\underline{\alpha}}$ is \emph{not} an $\mathrm{SL}(2)$ element, its determinant is related to $U_s{}^{\underline{s}}$ by
\be
\det ( U_\alpha{}^{\underline{\alpha}} ) = ( U_s{}^{\underline{s}} )^{-3/2} \,.
\ee
In order to simplify the notation, we will henceforth frequently drop the underlines from the indices (they can always be reintroduced by checking whether the matrix involved is $U$ or $U^{-1})$. We will also call $u \equiv U_{s}{}^{\underline{s}}$. 

It is straightforward to evaluate the components of the embedding tensor and trombone gauging from the general expressions in section \ref{SSgeneral}. One finds for the embedding tensor proper the components:
\be
\rho^{2\lambda} \Theta_{s \beta}{}^\alpha = - \frac{3}{4} \delta_\beta^\alpha \partial_s u^{-1} - u^{-1} U_\gamma{}^\alpha \partial_s \Uinv_\beta{}^\gamma \,,
\ee
\be
\rho^{2\lambda} \Theta_{\alpha s}{}^s = - \frac{3}{4} \partial_\gamma \Uinv_\alpha{}^\gamma 
- \Uinv_{\alpha}{}^\beta \partial_\beta \ln u^{-1} \,,
\ee
\be
\begin{split} 
\rho^{2\lambda} \Theta_{\beta \gamma}{}^{\alpha} & = 
+ U_\delta{}^\alpha \Uinv_{\gamma}{}^\epsilon \partial_\epsilon \Uinv_{\beta}{}^{\delta} 
- U_\delta{}^\alpha \Uinv_{\beta}{}^\epsilon \partial_\epsilon \Uinv_{\gamma}{}^{\delta} 
\\ & 
- \frac{1}{8} \delta_\beta^\alpha \partial_\delta \Uinv_\gamma{}^{\delta} 
+ \frac{1}{8} \delta_\gamma^\alpha \partial_\delta \Uinv_\beta{}^{\delta} \,.
\end{split} 
\ee
In fact, the latter two are not independent: one can show that
\be
\Theta_{\beta \gamma}{}^\alpha = 3 \delta^\alpha_{[\beta} \Theta_{\gamma ] s}{}^s \,.
\ee
We will denote $\Theta_{\alpha} \equiv \Theta_{\alpha s}{}^s$. 
One also has $\Theta_{s\gamma}{}^\gamma = 0$. Hence the embedding tensor components correspond to a $\mathbf{3}$ and $\mathbf{2}$ of $\mathrm{SL}(2)$. 

In addition, the trombone gaugings are 
\be
7 \rho^{2\lambda} \theta_\alpha  = \partial_\gamma \Uinv_\alpha{}^\gamma - 8 \Uinv_{\alpha}{}^\gamma \partial_\gamma \ln \rho^{2\lambda} \,,
\ee
\be
7 \rho^{2\lambda} \theta_s = \partial_s u^{-1}  - 8 u^{-1} \partial_s \ln \rho^{2\lambda} \,.
\ee
These give a further $\mathbf{2}$ and $\mathbf{1}$. 

The components of the generalised torsion built using the above are
\be
\rho^{2\lambda} \tau_{\alpha s}{}^s =  -\partial_\gamma \Uinv_\alpha{}^\gamma 
- \Uinv_{\alpha}{}^\beta \partial_\beta \ln u^{-1} 
+ 2 \Uinv_\alpha{}^\beta \partial_\beta \ln \rho^{2\lambda} \,,
\ee
\be
\rho^{2\lambda} \tau_{s \beta}{}^\alpha =  
-u^{-1} U_\gamma{}^\alpha \partial_s \Uinv_\beta{}^\gamma 
-\delta_\beta^\alpha \partial_s u^{-1} 
+2 \delta_\beta^\alpha u^{-1} \partial_s \ln \rho^{2\lambda} \,,
\ee
\be
\begin{split} 
\rho^{2\lambda} \tau_{\beta \gamma}{}^{\alpha} & = 
U_\delta{}^\alpha \Uinv_{\gamma}{}^\epsilon \partial_\epsilon \Uinv_{\beta}{}^{\delta} 
 -U_\delta{}^\alpha \Uinv_{\beta}{}^\epsilon \partial_\epsilon \Uinv_{\gamma}{}^{\delta} 
\\ & 
- \delta_\beta^\alpha \Uinv_\gamma{}^{\delta} \partial_\delta \ln \rho^{2\lambda} 
+ \delta_\gamma^\alpha \Uinv_\beta{}^{\delta} \partial_\delta \ln \rho^{2\lambda} \,.
\end{split} 
\ee
The fields strengths are deformed in the following manner:
\be
\mathcal{F}_{\mu \nu}{}^\alpha \rightarrow
\bar{\mathcal{F}}_{\mu \nu}{}^{\alpha} 
+ \tau_{\beta \gamma}{}^\alpha \bar A_\mu{}^\beta \bar A_\nu{}^\gamma 
+ \tau_{s \gamma}{}^\alpha \bar A_{[\mu}^{s} \bar A_{\nu]}{}^\gamma
+  \bar B_{\mu \nu}{}^{\beta s} \left( \frac{7}{4} \theta_s \delta_\beta^\alpha - \Theta_{s\beta}{}^{\alpha} \right) \,,
\label{eq:SSFa2}
\ee
\be
\mathcal{F}_{\mu \nu}{}^s \rightarrow
\bar{\mathcal{F}}_{\mu \nu}{}^{s} 
+  \tau_{\alpha s}{}^s \bar A_{[\mu}{}^\alpha \bar A_{\nu]}{}^s 
+\bar B_{\mu \nu}{}^{\alpha s} \left( \frac{7}{4} \theta_\alpha - \Theta_{\alpha}  \right) \,,
\label{eq:eqSSFa1} 
\ee
\be
\begin{split} 
\mathcal{H}_{\mu \nu \rho}{}^{\alpha s} & \rightarrow
\bar{\mathcal{H}}_{\mu \nu \rho}{}^{\alpha s}
+ 3
\tau_{\gamma \delta}{}^{\alpha} 
\bar A_{[\mu}{}^{\gamma} \bar B_{\nu \rho]}{}^{\delta s} 
+ 3
\tau_{s \gamma}{}^{\alpha} 
\bar A_{[\mu}{}^{s} \bar B_{\nu \rho]}{}^{\gamma s}
+ 3
\tau_{\gamma s}{}^s 
\bar A_{[\mu}{}^\gamma \bar B_{\nu \rho]}{}^{\alpha s} 
\\ & 
\quad - \tau_{\gamma s}{}^s \bar A_{[\mu}{}^\alpha \bar A_\nu{}^\gamma \bar A_{\rho]}{}^s 
-\tau_{\beta \gamma}{}^\alpha \bar{A}_{[\mu}{}^s \bar A_{\nu}{}^\beta \bar A_{\rho]}{}^\gamma
\\ & 
\quad + \bar C^{\beta \alpha s} \left( \frac{21}{8} \theta_\beta + \frac{1}{2} \Theta_{\beta}  \right) \,,
\end{split} 
\label{eq:SSFb} 
\ee
\be
\begin{split} 
\mathcal{J}_{\mu \nu \rho \sigma}{}^{\alpha \beta s} & \rightarrow 
\bar{\mathcal{J}}_{\mu \nu \rho \sigma}{}^{\alpha \beta s} 
+ 4\cdot 2 
\tau_{\gamma \delta}{}^{[\alpha|} 
\bar A_{[\mu}{}^{\gamma} \bar C_{\nu \rho \sigma]}{}^{\delta|\beta]s} 
+ 4\cdot 2
\tau_{s \gamma}{}^{[\alpha|} 
\bar A_{[\mu}{}^{s} \bar C_{\nu \rho \sigma]}{}^{\gamma |\beta ] s }
+ 4  \tau_{\gamma s}{}^s 
\bar A_{[\mu}{}^\gamma \bar C_{\nu \rho \sigma]}{}^{\alpha \beta s}
\\ & 
\quad 
- 6 \left( 
+ 2 \tau_{\gamma \delta}{}^{[\alpha|} \bar A_{[\mu}{}^\gamma \bar A_\nu{}^\delta 
 + 2\tau_{s \gamma}{}^{[\alpha|} \bar A_{[\mu}{}^s\bar A_\nu{}^\alpha
 + \bar B_{[\mu \nu|}{}^{\gamma s} \left[ 
 \frac{7}{4} \theta_s \delta^{[\alpha}_\beta 
 -\Theta_{s\gamma}{}^{[\alpha} \right] 
 \right) \bar B_{|\rho \sigma]}{}^{\beta ] s} 
\\ & 
\quad 
+ \bar D^{\alpha \beta ss} \frac{7}{2} \theta_s \,,
\end{split} 
\label{eq:SSFc} 
\ee
while we also have
\be
\begin{split} 
D_\mu \mathcal{M}_{\alpha \beta}  & \rightarrow 
\bar{D}_\mu \bar{\gM}_{\alpha \beta} 
- 2 \bar{A}_\mu{}^\delta \Theta_{\delta ( \alpha}{}^\gamma \bar{\gM}_{\beta) \gamma} 
- 2 \bar{A}_\mu{}^s \Theta_{s( \alpha}{}^\gamma \bar{\gM}_{\beta) \gamma} \\ 
&\qquad - \frac{7}{8} \left(
2 \theta_{(\alpha} \bar{A}_\mu{}^\gamma \bar{\gM}_{\beta)\gamma} 
- \frac{12}{7}  \theta_{s} \bar{A}_\mu{}^s \bar{\gM}_{\alpha \beta} 
+ \frac{2}{7}  \theta_{\gamma} \bar{A}_\mu{}^\gamma \bar{\gM}_{\alpha \beta} 
\right) \,,
\end{split} 
\label{eq:SSM22} 
\ee
\be
D_\mu \mathcal{M}_{ss} \rightarrow 
\bar{D}_\mu \bar{\gM}_{ss}
- 2 \bar A_\mu{}^\gamma \Theta_{\gamma}  \bar{\gM}_{ss} 
- \frac{7}{8} \left( 
\frac{16}{7} \theta_s \bar A_\mu{}^s \bar{\gM}_{ss} 
- \frac{12}{7} \theta_\gamma \bar A_\mu{}^\gamma \bar{\gM}_{ss}
\right) \,.
\label{eq:SSM11} 
\ee
If the trombone gaugings $\theta$ are zero, one can define an action for the reduced theory,
after integrating out the coordinates on which the twist matrices depend, i.e.
\be
S = \int \dd^9 X \dd^3Y \sqrt{g} \mathcal{L}(g,\gM,A, \dots ) 
= \int \dd^3 Y \rho^{-14\lambda} \dd^9X \sqrt{\bar{g}} \bar{\mathcal{L}}( \bar{g}, \bar{\gM}, \bar{A}, \dots ) \,.
\ee
The Lagrangian $\bar{\mathcal{L}}$ takes the same form as that of the original EFT, but with the field strengths modified as above and the scalar potential modified as follows:
The reduction of the scalar potential gives new terms involving the gaugings (up to total derivatives):
\be
\begin{split} 
V(\mathcal{M} ) & \rightarrow V(\bar{\gM}) \\ & + \bar{\mathcal{M}}^{ss} 
\Big(
- \frac{1}{2} \bar{\cH}^{\alpha \gamma} \bar{\cH}_{\beta \delta} \Theta_{s\alpha}{}^\beta \Theta_{s\gamma}{}^\delta
- \frac{1}{2} \Theta_{s\beta}{}^\alpha \Theta_{s\alpha}{}^\beta
- \Theta_{s\beta}{}^\alpha \bar{\cH}^{\beta \delta} \partial_s \bar{\cH}_{\alpha \delta} 
\Big) 
\\ &
+ \bar{\gM}_{ss}^{3/4} 
 \left( 
\frac{3}{2}  
\partial_{\alpha} \bar{\cH}^{\alpha \beta} 
\Theta_{\beta }
-
\frac{7}{8} \bar{\cH}^{\alpha \beta} \partial_{\alpha} \ln \bar{\gM}_{ss} \Theta_{\beta}
-2 
  \Theta_{\alpha} \Theta_{\beta} \right) \,. \end{split} 
\label{eq:SSV} 
\ee
If the trombone is non-zero, then one must work just with the equations of motion. 

\subsection{IIA section} 

We now describe some details of the dictionary relating the $\G$ EFT described above to (massive) IIA.

\subsubsection*{Ordinary IIA} 

The action of IIA supergravity in string frame is:
\be
\begin{split} 
 S_{IIA} 
= \int \dd^{10}X \sqrt{|\hat{g}|} & \Big(  e^{-2\Phi} \left[ R - \frac{1}{12} \hat H_{\hmu\hnu\hrho} \hat H^{\hmu\hnu\hrho} + 4 \partial_\hmu \Phi \partial^\hmu\Phi\right]
\\ & 
 - \frac{1}{4} \hat F_{\hmu\hnu} \hat F^{\hmu\hnu} - \frac{1}{48} \hat F_{\hmu\hnu\hrho\hlambda} \hat F^{\hmu\hnu\hrho\hlambda} 
+ \sqrt{|\hat{g}|}^{-1} \mathcal{L}_{CS} 
\Big) \,,
\end{split} 
\label{eq:IIAaction} 
\ee
with field strengths $\hat H_{\hmu\hnu\hrho} = 3 \partial_{[\hmu}\hat B_{\hnu\hrho]}$, $\hat F_{\hmu\hnu} = 2 \partial_{[\hmu} \hat C_{\hnu]}$ and $\hat{F}_{\hmu\hnu\hrho\hlambda} = 4 \partial_{[\hnu} \hat C_{\hnu\hrho\hlambda]} + 4 \hat C_{[\hmu} \hat H_{\hnu\hrho\hlambda]}$. In order to match with the $9+3$ split of the $\G$ EFT, we must impose a $9+1$ coordinate split such that $X^{\hmu}= ( X^\mu, X^9)$, where we will match $X^9 \equiv Y^2$. 

The 10-dimensional string frame metric $\hat g_{\hmu \hnu}$ is decomposed as in \eqref{gkk}
with $\Omega = \phi^{-1/7} e^{4\Phi/7}$ (where $\phi \equiv \hat g_{99}$), and the RR 1-form decomposed as \eqref{IIACdecomp}.
The EFT degrees of freedom can then be decomposed in terms of the fields of IIA supergravity. The generalised metric components encode the dilaton $\Phi$, metric scalar $\phi$ and one-form scalar $C_9$ as in
\eqref{IIAgm1} and \eqref{IIAgm2}. The components of the one-form $A_\mu{}^M$ are as in \eqref{IIAA}. The remaining form fields encoding the physical degrees of freedom are
\be
B_{\mu\nu}{}^{\alpha s} 
= 
\begin{pmatrix} 
C_{\mu\nu9} - C_{[\mu} B_{\nu]9} \\
- B_{\mu\nu} - A_{[\mu} B_{\nu]9} 
\end{pmatrix} \,,
\ee
and
\be
C_{\mu\nu\rho}{}^{12s} = C_{\mu\nu\rho} - 3 A_{[\mu} C_{\nu\rho]9} 
- 3 C_{[\mu} B_{\nu\rho]} 
- 4 C_{[\mu } A_{\nu} B_{\rho]9} \,.
\ee
The field strengths of the IIA supergravity appear in those of the $\G$ EFT via 
\be
\mathcal{F}_{\mu\nu}{}^\alpha = 
\begin{pmatrix} 
F_{\mu\nu} + 2 A_{[\mu}{} F_{\nu]9} - C_9 \mathcal{F}_{\mu\nu} 
\\
\mathcal{F}_{\mu\nu} 
\end{pmatrix} 
\quad , \quad 
\mathcal{F}_{\mu\nu}{}^s = - H_{\mu\nu9} \,, 
\ee
(here $\mathcal{F}_{\mu\nu} = 2 \partial_{[\mu} A_{\nu]} - 2 A_{[\mu} \partial_9 A_{\nu]}$ is the field strength of the Kaluza-Klein vector)
\be
\mathcal{H}_{\mu\nu\rho}{}^{\alpha s} = 
\begin{pmatrix} 
F_{\mu\nu\rho 9} + C_9 ( H_{\mu\nu\rho} - 3 A_{[\mu}{} H_{\nu\rho]9} )
\\ 
 - H_{\mu\nu\rho} + 3 A_{[\mu}{} H_{\nu\rho]9} 
\end{pmatrix} \,,
\ee
\be
\mathcal{J}_{\mu\nu\rho \sigma}{}^{12s} = F_{\mu\nu\rho \sigma} + 4 A_{[\mu}{} F_{\nu\rho \sigma]9} + 4 C_9 A_{[\mu}{} H_{\nu\rho \sigma]} \,,
\ee
as well as in 
\be
D_\mu \cH_{\alpha \beta} = 
\begin{pmatrix} 
D_\mu ( \phi^{-1/2} e^{\Phi} ) & \phi^{-1/2} e^\Phi F_{\mu 9} + C_9 D_\mu ( \phi^{-1/2} e^{\Phi} )\\
\phi^{-1/2} e^\Phi F_{\mu 9} + C_9  D_\mu ( \phi^{-1/2} e^{\Phi} ) & 
D_\mu ( \phi^{1/2} e^{-\Phi} ) + 2 C_9 \phi^{-1/2} e^\Phi F_{\mu 9 } 
+(C_9)^2 D_\mu ( \phi^{-1/2} e^{\Phi} )
\end{pmatrix} \,.
\ee

\subsubsection*{Massive IIA}

We give here the full deformations relevant to checking how the field strengths are modified. 
The twist matrix is
\be
U_{\alpha}{}^\beta ( Y^M ) = \begin{pmatrix} 1 & 0 \\ my^s & 1 \end{pmatrix}  \quad, \quad U_s{}^s (Y^M) = 1 \quad , \quad \rho(Y^M) = 1 \,.
\ee
The non-trivial twistings of the field strengths in the tensor hierarchy are, using \eqref{eq:SSFa2} to \eqref{eq:SSFc},
\be
\mathcal{F}_{\mu \nu}{}^1 \rightarrow
\bar{\mathcal{F}}_{\mu \nu}{}^1 + m \bar A_{[\mu}{}^s\bar A_{\nu]}{}^2 
- m\bar B_{\mu \nu}{}^{2s} \,,
\ee
\be
\mathcal{H}_{\mu \nu \rho}{}^{1 s} \rightarrow \bar{\mathcal{H}}_{\mu \nu \rho}{}^{1s} + 3m \bar A_{[\mu}{}^s  \bar B_{\nu \rho]}{}^{2s} \,,
\ee
\be
\mathcal{J}_{\mu\nu\rho\sigma}{}^{12s} \rightarrow 
\bar{\mathcal{J}}_{\mu\nu\rho\sigma}{}^{12s} 
+ 3 m \bar B_{[\mu \nu}{}^{2s} \bar B_{\rho\sigma]}{}^{2s} 
-6m \bar A_{[\mu}{}^s \bar A_{\nu}{}^2 \bar B_{\rho\sigma]}{}^{2s} \,,
\ee
while one also has from \eqref{eq:SSM22} and \eqref{eq:SSM11} 
\be
D_\mu \cH_{12} \rightarrow D_\mu \bar{\cH}_{12} - m \bar A_\mu{}^s \bar{\cH}_{11}
\quad,\quad
D_\mu \cH_{22} \rightarrow D_\mu \bar{\cH}_{22} - 2 m \bar A_\mu{}^s \bar{\cH}_{21}\,.
\ee

\bibliography{NewBib}

\providecommand{\href}[2]{#2}\begingroup\raggedright\begin{thebibliography}{10}

\bibitem{deAzcarraga:1991px}
J.~A. de~Azcarraga, J.~M. Izquierdo, and P.~K. Townsend, {\it {A Kaluza-Klein
  origin for the superstring tension}},  {\em Phys. Rev.} {\bf D45} (1992)
  R3321--R3325.

\bibitem{Townsend:1992fa}
P.~K. Townsend, {\it {World sheet electromagnetism and the superstring
  tension}},  {\em Phys. Lett.} {\bf B277} (1992) 285--288.

\bibitem{Bergshoeff:1992gq}
E.~Bergshoeff, L.~A.~J. London, and P.~K. Townsend, {\it {Space-time scale
  invariance and the superp-brane}},  {\em Class. Quant. Grav.} {\bf 9} (1992)
  2545--2556, [\href{http://arxiv.org/abs/hep-th/9206026}{{\tt
  hep-th/9206026}}].

\bibitem{Townsend:1997kr}
P.~K. Townsend, {\it {Membrane tension and manifest IIB S duality}},  {\em
  Phys. Lett.} {\bf B409} (1997) 131--135,
  [\href{http://arxiv.org/abs/hep-th/9705160}{{\tt hep-th/9705160}}].

\bibitem{Cederwall:1997ts}
M.~Cederwall and P.~K. Townsend, {\it {The Manifestly Sl(2,Z) covariant
  superstring}},  {\em JHEP} {\bf 09} (1997) 003,
  [\href{http://arxiv.org/abs/hep-th/9709002}{{\tt hep-th/9709002}}].

\bibitem{Bergshoeff:2003sy}
E.~Bergshoeff, U.~Gran, R.~Linares, M.~Nielsen, and D.~Roest, {\it {Domain
  walls and the creation of strings}},  {\em Class. Quant. Grav.} {\bf 20}
  (2003) 3465--3482, [\href{http://arxiv.org/abs/hep-th/0303253}{{\tt
  hep-th/0303253}}].

\bibitem{Bergshoeff:2006gs}
E.~A. Bergshoeff, M.~de~Roo, S.~F. Kerstan, T.~Ortin, and F.~Riccioni, {\it
  {SL(2,R)-invariant IIB Brane Actions}},  {\em JHEP} {\bf 02} (2007) 007,
  [\href{http://arxiv.org/abs/hep-th/0611036}{{\tt hep-th/0611036}}].

\bibitem{Siegel:1993th}
W.~Siegel, {\it {Superspace duality in low-energy superstrings}},  {\em
  Phys.Rev.} {\bf D48} (1993) 2826--2837,
  [\href{http://arxiv.org/abs/hep-th/9305073}{{\tt hep-th/9305073}}].

\bibitem{Siegel:1993xq}
W.~Siegel, {\it {Two vierbein formalism for string inspired axionic gravity}},
  {\em Phys.Rev.} {\bf D47} (1993) 5453--5459,
  [\href{http://arxiv.org/abs/hep-th/9302036}{{\tt hep-th/9302036}}].

\bibitem{Hull:2009mi}
C.~Hull and B.~Zwiebach, {\it {Double Field Theory}},  {\em JHEP} {\bf 0909}
  (2009) 099, [\href{http://arxiv.org/abs/0904.4664}{{\tt arXiv:0904.4664}}].

\bibitem{Hull:2009zb}
C.~Hull and B.~Zwiebach, {\it {The Gauge algebra of double field theory and
  Courant brackets}},  {\em JHEP} {\bf 0909} (2009) 090,
  [\href{http://arxiv.org/abs/0908.1792}{{\tt arXiv:0908.1792}}].

\bibitem{Hohm:2010jy}
O.~Hohm, C.~Hull, and B.~Zwiebach, {\it {Background independent action for
  double field theory}},  {\em JHEP} {\bf 1007} (2010) 016,
  [\href{http://arxiv.org/abs/1003.5027}{{\tt arXiv:1003.5027}}].

\bibitem{Hohm:2010pp}
O.~Hohm, C.~Hull, and B.~Zwiebach, {\it {Generalized metric formulation of
  double field theory}},  {\em JHEP} {\bf 1008} (2010) 008,
  [\href{http://arxiv.org/abs/1006.4823}{{\tt arXiv:1006.4823}}].

\bibitem{Berman:2010is}
D.~S. Berman and M.~J. Perry, {\it {Generalized Geometry and M theory}},  {\em
  JHEP} {\bf 1106} (2011) 074, [\href{http://arxiv.org/abs/1008.1763}{{\tt
  arXiv:1008.1763}}].

\bibitem{Berman:2011jh}
D.~S. Berman, H.~Godazgar, M.~J. Perry, and P.~West, {\it {Duality Invariant
  Actions and Generalised Geometry}},  {\em JHEP} {\bf 1202} (2012) 108,
  [\href{http://arxiv.org/abs/1111.0459}{{\tt arXiv:1111.0459}}].

\bibitem{Berman:2011cg}
D.~S. Berman, H.~Godazgar, M.~Godazgar, and M.~J. Perry, {\it {The Local
  symmetries of M-theory and their formulation in generalised geometry}},  {\em
  JHEP} {\bf 1201} (2012) 012, [\href{http://arxiv.org/abs/1110.3930}{{\tt
  arXiv:1110.3930}}].

\bibitem{Berman:2012vc}
D.~S. Berman, M.~Cederwall, A.~Kleinschmidt, and D.~C. Thompson, {\it {The
  gauge structure of generalised diffeomorphisms}},  {\em JHEP} {\bf 1301}
  (2013) 064, [\href{http://arxiv.org/abs/1208.5884}{{\tt arXiv:1208.5884}}].

\bibitem{Berman:2015rcc}
D.~S. Berman, C.~D.~A. Blair, E.~Malek, and F.~J. Rudolph, {\it {An Action for
  F-theory: $\mathrm{SL}(2) \times \mathbb{R}^+$ Exceptional Field Theory}},
  \href{http://arxiv.org/abs/1512.06115}{{\tt arXiv:1512.06115}}.

\bibitem{Hohm:2015xna}
O.~Hohm and Y.-N. Wang, {\it {Tensor Hierarchy and Generalized Cartan Calculus
  in SL(3)$\times$SL(2) Exceptional Field Theory}},
  \href{http://arxiv.org/abs/1501.01600}{{\tt arXiv:1501.01600}}.

\bibitem{Musaev:2015ces}
E.~T. Musaev, {\it {Exceptional field theory: $SL(5)$}},  {\em JHEP} {\bf 02}
  (2016) 012, [\href{http://arxiv.org/abs/1512.02163}{{\tt arXiv:1512.02163}}].

\bibitem{Abzalov:2015ega}
A.~Abzalov, I.~Bakhmatov, and E.~T. Musaev, {\it {Exceptional field theory:
  $SO(5,5)$}},  {\em JHEP} {\bf 06} (2015) 088,
  [\href{http://arxiv.org/abs/1504.01523}{{\tt arXiv:1504.01523}}].

\bibitem{Hohm:2013vpa}
O.~Hohm and H.~Samtleben, {\it {Exceptional Field Theory I: $E_{6(6)}$
  covariant Form of M-Theory and Type IIB}},  {\em Phys.Rev.} {\bf D89} (2014)
  066016, [\href{http://arxiv.org/abs/1312.0614}{{\tt arXiv:1312.0614}}].

\bibitem{Hohm:2013uia}
O.~Hohm and H.~Samtleben, {\it {Exceptional Field Theory II: E$_{7(7)}$}},
  {\em Phys.Rev.} {\bf D89} (2014) 066017,
  [\href{http://arxiv.org/abs/1312.4542}{{\tt arXiv:1312.4542}}].

\bibitem{Hohm:2014fxa}
O.~Hohm and H.~Samtleben, {\it {Exceptional Field Theory III: E$_{8(8)}$}},
  {\em Phys.Rev.} {\bf D90} (2014) 066002,
  [\href{http://arxiv.org/abs/1406.3348}{{\tt arXiv:1406.3348}}].

\bibitem{Godazgar:2014nqa}
H.~Godazgar, M.~Godazgar, O.~Hohm, H.~Nicolai, and H.~Samtleben, {\it
  {Supersymmetric E$_{7(7)}$ Exceptional Field Theory}},  {\em JHEP} {\bf 09}
  (2014) 044, [\href{http://arxiv.org/abs/1406.3235}{{\tt arXiv:1406.3235}}].

\bibitem{Musaev:2014lna}
E.~Musaev and H.~Samtleben, {\it {Fermions and supersymmetry in E$_{6(6)}$
  exceptional field theory}},  {\em JHEP} {\bf 03} (2015) 027,
  [\href{http://arxiv.org/abs/1412.7286}{{\tt arXiv:1412.7286}}].

\bibitem{Hull:2004in}
C.~Hull, {\it {A Geometry for non-geometric string backgrounds}},  {\em JHEP}
  {\bf 0510} (2005) 065, [\href{http://arxiv.org/abs/hep-th/0406102}{{\tt
  hep-th/0406102}}].

\bibitem{Hull:2006va}
C.~M. Hull, {\it {Doubled Geometry and T-Folds}},  {\em JHEP} {\bf 0707} (2007)
  080, [\href{http://arxiv.org/abs/hep-th/0605149}{{\tt hep-th/0605149}}].

\bibitem{Rocek:1991ps}
M.~Rocek and E.~P. Verlinde, {\it {Duality, quotients, and currents}},  {\em
  Nucl.Phys.} {\bf B373} (1992) 630--646,
  [\href{http://arxiv.org/abs/hep-th/9110053}{{\tt hep-th/9110053}}].

\bibitem{Park:2013mpa}
J.-H. Park, {\it {Comments on double field theory and diffeomorphisms}},  {\em
  JHEP} {\bf 1306} (2013) 098, [\href{http://arxiv.org/abs/1304.5946}{{\tt
  arXiv:1304.5946}}].

\bibitem{Lee:2013hma}
K.~Lee and J.-H. Park, {\it {Covariant action for a string in doubled yet
  gauged spacetime}},  \href{http://arxiv.org/abs/1307.8377}{{\tt
  arXiv:1307.8377}}.

\bibitem{Bergshoeff:1997gy}
E.~Bergshoeff, B.~Janssen, and T.~Ortin, {\it {Kaluza-Klein monopoles and
  gauged sigma models}},  {\em Phys. Lett.} {\bf B410} (1997) 131--141,
  [\href{http://arxiv.org/abs/hep-th/9706117}{{\tt hep-th/9706117}}].

\bibitem{Romans:1985tz}
L.~J. Romans, {\it {Massive N=2a Supergravity in Ten-Dimensions}},  {\em Phys.
  Lett.} {\bf B169} (1986) 374.

\bibitem{Hohm:2011cp}
O.~Hohm and S.~K. Kwak, {\it {Massive Type II in Double Field Theory}},  {\em
  JHEP} {\bf 1111} (2011) 086, [\href{http://arxiv.org/abs/1108.4937}{{\tt
  arXiv:1108.4937}}].

\bibitem{Ciceri:2016dmd}
F.~Ciceri, A.~Guarino, and G.~Inverso, {\it {The exceptional story of massive
  IIA supergravity}},  {\em JHEP} {\bf 08} (2016) 154,
  [\href{http://arxiv.org/abs/1604.08602}{{\tt arXiv:1604.08602}}].

\bibitem{Geissbuhler:2011mx}
D.~Geissb{\"u}hler, {\it {Double Field Theory and N=4 Gauged Supergravity}},
  {\em JHEP} {\bf 1111} (2011) 116, [\href{http://arxiv.org/abs/1109.4280}{{\tt
  arXiv:1109.4280}}].

\bibitem{Aldazabal:2011nj}
G.~Aldazabal, W.~Baron, D.~Marqu{\'e}s, and C.~N{\'u}{\~n}ez, {\it {The
  effective action of Double Field Theory}},  {\em JHEP} {\bf 1111} (2011) 052,
  [\href{http://arxiv.org/abs/1109.0290}{{\tt arXiv:1109.0290}}].

\bibitem{Grana:2012rr}
M.~Gra{\~n}a and D.~Marqu{\'e}s, {\it {Gauged Double Field Theory}},  {\em
  JHEP} {\bf 1204} (2012) 020, [\href{http://arxiv.org/abs/1201.2924}{{\tt
  arXiv:1201.2924}}].

\bibitem{Berman:2012uy}
D.~S. Berman, E.~T. Musaev, and D.~C. Thompson, {\it {Duality Invariant
  M-theory: Gauged supergravities and Scherk-Schwarz reductions}},  {\em JHEP}
  {\bf 1210} (2012) 174, [\href{http://arxiv.org/abs/1208.0020}{{\tt
  arXiv:1208.0020}}].

\bibitem{Hohm:2014qga}
O.~Hohm and H.~Samtleben, {\it {Consistent Kaluza-Klein Truncations via
  Exceptional Field Theory}},  {\em JHEP} {\bf 01} (2015) 131,
  [\href{http://arxiv.org/abs/1410.8145}{{\tt arXiv:1410.8145}}].

\bibitem{Gualtieri:2003dx}
M.~Gualtieri, {\it {Generalized complex geometry}},
  \href{http://arxiv.org/abs/math/0401221}{{\tt math/0401221}}.

\bibitem{Hull:2007zu}
C.~Hull, {\it {Generalised Geometry for M-Theory}},  {\em JHEP} {\bf 0707}
  (2007) 079, [\href{http://arxiv.org/abs/hep-th/0701203}{{\tt
  hep-th/0701203}}].

\bibitem{Coimbra:2011nw}
A.~Coimbra, C.~Strickland-Constable, and D.~Waldram, {\it {Supergravity as
  Generalised Geometry I: Type II Theories}},  {\em JHEP} {\bf 1111} (2011)
  091, [\href{http://arxiv.org/abs/1107.1733}{{\tt arXiv:1107.1733}}].

\bibitem{Coimbra:2011ky}
A.~Coimbra, C.~Strickland-Constable, and D.~Waldram, {\it {$E_{d(d)} \times
  \mathbb{R}^+$ generalised geometry, connections and M theory}},  {\em JHEP}
  {\bf 1402} (2014) 054, [\href{http://arxiv.org/abs/1112.3989}{{\tt
  arXiv:1112.3989}}].

\bibitem{Coimbra:2012af}
A.~Coimbra, C.~Strickland-Constable, and D.~Waldram, {\it {Supergravity as
  Generalised Geometry II: $E_{d(d)} \times \mathbb{R}^+$ and M theory}},  {\em
  JHEP} {\bf 1403} (2014) 019, [\href{http://arxiv.org/abs/1212.1586}{{\tt
  arXiv:1212.1586}}].

\bibitem{Cassani:2016ncu}
D.~Cassani, O.~de~Felice, M.~Petrini, C.~Strickland-Constable, and D.~Waldram,
  {\it {Exceptional generalised geometry for massive IIA and consistent
  reductions}},  {\em JHEP} {\bf 08} (2016) 074,
  [\href{http://arxiv.org/abs/1605.00563}{{\tt arXiv:1605.00563}}].

\bibitem{Bergshoeff:1997ak}
E.~Bergshoeff, Y.~Lozano, and T.~Ortin, {\it {Massive branes}},  {\em Nucl.
  Phys.} {\bf B518} (1998) 363--423,
  [\href{http://arxiv.org/abs/hep-th/9712115}{{\tt hep-th/9712115}}].

\bibitem{Duff:1989tf}
M.~Duff, {\it {Duality rotations in string theory}},  {\em Nucl.Phys.} {\bf
  B335} (1990) 610.

\bibitem{Duff:1990hn}
M.~Duff and J.~Lu, {\it {Duality Rotations in Membrane Theory}},  {\em
  Nucl.Phys.} {\bf B347} (1990) 394--419.

\bibitem{Tseytlin:1990nb}
A.~A. Tseytlin, {\it {Duality symmetric formulation of string world sheet
  dynamics}},  {\em Phys.Lett.} {\bf B242} (1990) 163--174.

\bibitem{Tseytlin:1990va}
A.~A. Tseytlin, {\it {Duality symmetric closed string theory and interacting
  chiral scalars}},  {\em Nucl.Phys.} {\bf B350} (1991) 395--440.

\bibitem{Duff:2015jka}
M.~J. Duff, J.~X. Lu, R.~Percacci, C.~N. Pope, H.~Samtleben, and E.~Sezgin,
  {\it {Membrane Duality Revisited}},  {\em Nucl. Phys.} {\bf B901} (2015)
  1--21, [\href{http://arxiv.org/abs/1509.02915}{{\tt arXiv:1509.02915}}].

\bibitem{Sakatani:2016sko}
Y.~Sakatani and S.~Uehara, {\it {Branes in Extended Spacetime: Brane
  Worldvolume Theory Based on Duality Symmetry}},  {\em Phys. Rev. Lett.} {\bf
  117} (2016), no.~19 191601, [\href{http://arxiv.org/abs/1607.04265}{{\tt
  arXiv:1607.04265}}].

\bibitem{Bandos:2015rvs}
I.~Bandos, {\it {On section conditions of E$_{7(+7)}$ exceptional field theory
  and superparticle in $ \mathcal{N}=8 $ central charge superspace}},  {\em
  JHEP} {\bf 01} (2016) 132, [\href{http://arxiv.org/abs/1512.02287}{{\tt
  arXiv:1512.02287}}].

\bibitem{Bandos:2016ppv}
I.~Bandos, {\it {Exceptional field theories, superparticles in an enlarged 11D
  superspace and higher spin theories}},
  \href{http://arxiv.org/abs/1612.01321}{{\tt arXiv:1612.01321}}.

\bibitem{Berkeley:2014nza}
J.~Berkeley, D.~S. Berman, and F.~J. Rudolph, {\it {Strings and Branes are
  Waves}},  {\em JHEP} {\bf 06} (2014) 006,
  [\href{http://arxiv.org/abs/1403.7198}{{\tt arXiv:1403.7198}}].

\bibitem{Berman:2014jsa}
D.~S. Berman and F.~J. Rudolph, {\it {Branes are Waves and Monopoles}},  {\em
  JHEP} {\bf 05} (2015) 015, [\href{http://arxiv.org/abs/1409.6314}{{\tt
  arXiv:1409.6314}}].

\bibitem{Berman:2014hna}
D.~S. Berman and F.~J. Rudolph, {\it {Strings, Branes and the Self-dual
  Solutions of Exceptional Field Theory}},  {\em JHEP} {\bf 05} (2015) 130,
  [\href{http://arxiv.org/abs/1412.2768}{{\tt arXiv:1412.2768}}].

\bibitem{Blair:2015eba}
C.~D.~A. Blair, {\it {Conserved Currents of Double Field Theory}},  {\em JHEP}
  {\bf 04} (2016) 180, [\href{http://arxiv.org/abs/1507.07541}{{\tt
  arXiv:1507.07541}}].

\bibitem{Park:2015bza}
J.-H. Park, S.-J. Rey, W.~Rim, and Y.~Sakatani, {\it {O(D, D) covariant Noether
  currents and global charges in double field theory}},  {\em JHEP} {\bf 11}
  (2015) 131, [\href{http://arxiv.org/abs/1507.07545}{{\tt arXiv:1507.07545}}].

\bibitem{Naseer:2015fba}
U.~Naseer, {\it {Canonical formulation and conserved charges of double field
  theory}},  {\em JHEP} {\bf 10} (2015) 158,
  [\href{http://arxiv.org/abs/1508.00844}{{\tt arXiv:1508.00844}}].

\bibitem{Hohm:2013nja}
O.~Hohm and H.~Samtleben, {\it {Gauge theory of Kaluza-Klein and winding
  modes}},  {\em Phys.Rev.} {\bf D88} (2013) 085005,
  [\href{http://arxiv.org/abs/1307.0039}{{\tt arXiv:1307.0039}}].

\bibitem{Wang:2015hca}
Y.-N. Wang, {\it {Generalized Cartan Calculus in general dimension}},  {\em
  JHEP} {\bf 07} (2015) 114, [\href{http://arxiv.org/abs/1504.04780}{{\tt
  arXiv:1504.04780}}].

\bibitem{Morand:2017fnv}
K.~Morand and J.-H. Park, {\it {Classification of non-Riemannian
  doubled-yet-gauged spacetime}},  \href{http://arxiv.org/abs/1707.03713}{{\tt
  arXiv:1707.03713}}.

\bibitem{Blair:2013gqa}
C.~D.~A. Blair, E.~Malek, and J.-H. Park, {\it {M-theory and Type IIB from a
  Duality Manifest Action}},  {\em JHEP} {\bf 1401} (2014) 172,
  [\href{http://arxiv.org/abs/1311.5109}{{\tt arXiv:1311.5109}}].

\bibitem{Cederwall:2013naa}
M.~Cederwall, J.~Edlund, and A.~Karlsson, {\it {Exceptional geometry and tensor
  fields}},  {\em JHEP} {\bf 1307} (2013) 028,
  [\href{http://arxiv.org/abs/1302.6736}{{\tt arXiv:1302.6736}}].

\bibitem{Ko:2016dxa}
S.~M. Ko, J.-H. Park, and M.~Suh, {\it {The rotation curve of a point particle
  in stringy gravity}},  {\em JCAP} {\bf 1706} (2017), no.~06 002,
  [\href{http://arxiv.org/abs/1606.09307}{{\tt arXiv:1606.09307}}].

\bibitem{CederwallKorea}
M.~Cederwall, {\it {(Brane) Charges for 1/2 BPS in Exceptional Geometry}},
  presented at the workshop ``Duality and Novel Geometry in M-theory'', Asia
  Pacific Centre for Theoretical Physics, Postech, 2016.

\bibitem{MalekTalk}
E.~Malek, {\it {Membranes in Exceptional Generalised Geometry / EFT}},
  presented at the workshop ``Generalized Geometry \& T-Duality'', Simons
  Centre for Geometry and Physics, SUNY Stony Brook, 2016.

\bibitem{BCM}
D.~S. Berman, M.~Cederwall, and E.~Malek {\em unpublished}.

\bibitem{AbouZeid:1999fv}
M.~Abou-Zeid, B.~de~Wit, D.~Lust, and H.~Nicolai, {\it {Space-time
  supersymmetry, IIA / B duality and M theory}},  {\em Phys. Lett.} {\bf B466}
  (1999) 144--152, [\href{http://arxiv.org/abs/hep-th/9908169}{{\tt
  hep-th/9908169}}].

\bibitem{Bergshoeff:1996ui}
E.~Bergshoeff, M.~de~Roo, M.~B. Green, G.~Papadopoulos, and P.~K. Townsend,
  {\it {Duality of type II 7 branes and 8 branes}},  {\em Nucl. Phys.} {\bf
  B470} (1996) 113--135, [\href{http://arxiv.org/abs/hep-th/9601150}{{\tt
  hep-th/9601150}}].

\bibitem{Hull:1998vy}
C.~Hull, {\it {Massive string theories from M theory and F theory}},  {\em
  JHEP} {\bf 9811} (1998) 027, [\href{http://arxiv.org/abs/hep-th/9811021}{{\tt
  hep-th/9811021}}].

\bibitem{BBMR}
D.~S. Berman, C.~D.~A. Blair, E.~Malek, and F.~J. Rudolph {\em in progress}.

\bibitem{Malek:2016vsh}
E.~Malek, {\it {From Exceptional Field Theory to Heterotic Double Field Theory
  via K3}},  {\em JHEP} {\bf 03} (2017) 057,
  [\href{http://arxiv.org/abs/1612.01990}{{\tt arXiv:1612.01990}}].

\bibitem{deBoer:2012ma}
J.~de~Boer and M.~Shigemori, {\it {Exotic Branes in String Theory}},  {\em
  Phys.Rept.} {\bf 532} (2013) 65--118,
  [\href{http://arxiv.org/abs/1209.6056}{{\tt arXiv:1209.6056}}].

\bibitem{Blair:2016xnn}
C.~D.~A. Blair, {\it {Doubled strings, negative strings and null waves}},  {\em
  JHEP} {\bf 11} (2016) 042, [\href{http://arxiv.org/abs/1608.06818}{{\tt
  arXiv:1608.06818}}].

\bibitem{Blair:2013noa}
C.~D.~A. Blair, E.~Malek, and A.~J. Routh, {\it {An $O(D, D)$ invariant
  Hamiltonian action for the superstring}},  {\em Class.Quant.Grav.} {\bf 31}
  (2014), no.~20 205011, [\href{http://arxiv.org/abs/1308.4829}{{\tt
  arXiv:1308.4829}}].

\bibitem{Bandos:2015cha}
I.~Bandos, {\it {Superstring in doubled superspace}},  {\em Phys. Lett.} {\bf
  B751} (2015) 408--412, [\href{http://arxiv.org/abs/1507.07779}{{\tt
  arXiv:1507.07779}}].

\bibitem{Driezen:2016tnz}
S.~Driezen, A.~Sevrin, and D.~C. Thompson, {\it {Aspects of the Doubled
  Worldsheet}},  {\em JHEP} {\bf 12} (2016) 082,
  [\href{http://arxiv.org/abs/1609.03315}{{\tt arXiv:1609.03315}}].

\bibitem{Park:2016sbw}
J.-H. Park, {\it {Green-Schwarz superstring on doubled-yet-gauged spacetime}},
  {\em JHEP} {\bf 11} (2016) 005, [\href{http://arxiv.org/abs/1609.04265}{{\tt
  arXiv:1609.04265}}].

\end{thebibliography}\endgroup

\end{document}